# Intertwined orders from symmetry projected wavefunctions of repulsively interacting Fermi gases in optical lattices

A. Leprévost[1], O. Juillet[1] and R. Frésard[2]

[1] *Laboratoire LPC Caen, ENSICAEN, Université de Caen, CNRS/IN2P3, Caen, France.*

[2] *Laboratoire CRISMAT, UMR CNRS-ENSICAEN 6508, Caen, France.*

---

Unconventional strongly correlated phases of the repulsive Fermi-Hubbard model, which could be emulated by ultracold vapors loaded in optical lattices, are investigated by means of energy minimizations with quantum number projection before variation and without any assumed order parameter. In a tube-like geometry of optical plaquettes to realize the four-leg ladder Hubbard Hamiltonian, we highlight the intertwining of spin-, charge-, and pair-density waves embedded in a uniform *d*-wave superfluid background. As the lattice filling increases, this phase emerges from homogenous states exhibiting spiral magnetism and evolves towards a doped antiferromagnet. A concomitant enhancement of long-ranged *d*-wave pairing correlations is also found. Numerical tests of the approach for two-dimensional clusters are carried out, too.

---

## 1. Introduction

Low dimensional interacting quantum matter generally exhibits several phases at low energy that challenges the ability to distinguish between competing orders and their intertwining within one single correlated state [1]. Ultracold atoms provide an ideal playground to capture the essence of this problematic by their potential to properly emulate the fundamental mechanisms of quantum many-body physics [2]. In the fermionic sector, the BCS-to-BEC crossover [3,4] and the question of Stoner's itinerant ferromagnetism in repulsive gases [5,6] have been investigated. By trapping atomic vapors in optical lattices, a mimic of ideal crystalline matter can also be achieved [7]. By now, direct images of Fermi surfaces in the non-interacting limit [8] as well as *s*-wave superfluidity near unitary scattering [9] have been reported. Away from a Feshbach resonance, one is able to engineer almost perfectly the celebrated Hubbard model that has been first considered to describe the magnetism of metallic systems [10]. More generally, it aims to grasp the generic properties of spin-$1/2$ fermions moving on a lattice by hopping between neighboring sites $\langle \vec{r}, \vec{r}' \rangle$ and experiencing a local two-body interaction of strength $U$. In second-quantized form, the Hamiltonian is given by

$$\hat{H} = -t \sum_{\langle \vec{r}, \vec{r}' \rangle \sigma} \hat{c}^{\dagger}_{\vec{r}\sigma} \hat{c}_{\vec{r}'\sigma} + U \sum_{\vec{r}} \hat{n}_{\vec{r}\uparrow} \hat{n}_{\vec{r}\downarrow} , \qquad (1)$$

with $t$ the hopping integral; The fermionic creation, annihilation and density operators at site $\vec{r}$ with spin label $\sigma \in \{\uparrow, \downarrow\}$ are $\hat{c}^{\dagger}_{\vec{r}\sigma}$, $\hat{c}_{\vec{r}\sigma}$ and $\hat{n}_{\vec{r}\sigma} = \hat{c}^{\dagger}_{\vec{r}\sigma} \hat{c}_{\vec{r}\sigma}$, respectively. In the attractive regime, spin-polarized systems could exhibit several exotic superfluid phases [11] while the BCS-to-BEC transition has been addressed in the spin-balanced model [12]. Otherwise, the on-site repulsion can stand for a perfectly screened Coulomb interaction and it received a considerable renewed interest in two-dimensional (2D) geometry after Anderson's proposal [13] in connection to the spectacular properties of the high-$T_c$ cuprates. However, there is still no consensus about the adequacy of the positive-$U$ Hubbard model to capture the interplay between $d$-wave superconductivity, magnetism and inhomogeneous phases of copper oxides. This challenging issue is even more relevant since latest condensed-matter experiments seem to be consistent with an intriguing scenario where spin, density and long-ranged pair correlations develop cooperatively and are spatially modulated [14,15].

The exact answer to the question of whether the 2D repulsive Hubbard model supports such intertwining of multiple orders will probably be provided only through quantum emulators like ultracold atoms. Indeed, exact low energy properties of the Hamiltonian Eq. (1) are only accessible in one dimension [16] and for the infinitely connected Bethe lattice through the dynamical mean-field theory [17]. In other cases, computational methods to recover exact ground-states are generally marred by exponential complexity [18,19]. Nevertheless, diagrammatic quantum Monte-Carlo (QMC) simulations in continuous-time have recently allowed for a determination of the phase diagram at weak coupling for small to intermediate filling [20]. Even if ultracold fermions in optical lattices already enabled to monitor the Mott transition [21] and the development of antiferromagnetic correlations at half-filling [22], the knowledge of the phase diagram at low temperature and up to the strongly repulsive limit remains a long-term goal. In spirit of the compelling example provided by unitary Fermi gases [23], it is highly desirable to introduce theoretical approximate schemes that could guide experiments and benefit from the progressive results of this emulation. In order to embrace the full complexity of the repulsive Hubbard model, we set up in this paper a variational approach where the ground-state is progressively reconstructed from an expansion on symmetry adapted wavefunctions without any a priori input on the relevant correlations. The key features of the method are presented in Section 2. Its reliability against other numerical simulations is discussed in Section 3. Finally, we proceed in Section 4 to a systematic application in a four-leg ladder geometry motivated by recent experimental achievement of optical lattice plaquettes [24]. The obtained quantum phase diagram in the lattice filling-interaction strength plane highlights the intertwining of magnetic, density and pairing channels.

## 2. Methodology: The symmetry projected Hartree-Fock/Bogoliubov-de Gennes scheme

For weak coupling strength $U/t$, the determination of correlations that spontaneously emerge from the Hubbard Hamiltonian Eq. (1) can be achieved by identifying the channels in which instabilities develop through self-consistent perturbative or functional renormalization group methods [25,26]. In the strongly correlated regime, the problem could ideally be tackled with Gutzwiller-type wavefunctions $|\Psi_g\rangle = \hat{P}_G|\Phi\rangle$ where the operator $\hat{P}_G = \prod_{\vec{r}}\left(\hat{1} - g\hat{n}_{\vec{r}\uparrow}\hat{n}_{\vec{r}\downarrow}\right)$ partially suppresses the double occupancy entailed in a mean-field state $|\Phi\rangle$ through the real parameter $g$ [27]. Yet, the energy minimization has to be performed in a variational Monte-Carlo framework, rendering unrestricted calculations beyond reach. Hence, the reference wavefunction must be parameterized with a limited number of relevant variables to describe specific phases, such as $d$-wave superfluids [28], spirals [29] or stripes [30]. A step towards unbiased Gutzwiller calculations has been recently achieved [31]. However, orders exhibiting a periodicity larger than a few lattice spacings were forbidden, in contradiction to approximate QMC results [32] revealing long wavelength modes in ground-states.

Alternatively, correlations beyond mean-field can be generated by restoring deliberately broken symmetries through quantum number projection. In fact, the Hamiltonian Eq. (1) is invariant under local $U(1)$ gauge transformations, lattice translations, spin rotations and discrete symmetries of the lattice. Thus, exact eigenstates are characterized by the number of fermions $N$, the total pseudo-momentum $\vec{K}$, the total spin $S$ and its *z*-component $S_z$, as well as an irreducible representation of the lattice symmetry group. All these labels will be collectively denoted by $\Gamma$ in the following. Their restoration on top of a single Hartree-Fock (HF) wavefunction and before energy minimization recently yielded encouraging results for 2D clusters [33]. In particular, the exact ground-state of the four-site model has been analytically recovered irrespective of the interaction strength [34]. The approach, and its analog with several Slater determinants [35,36], also proved capable to evidence interplay between spin, charge and pair degrees of freedom. Potential superfluid features would nevertheless require a very large number of Hartree-Fock (HF) basis states to be accurately captured, whereas Bogoliubov-de Gennes (BdG) ansätze are well known to be more appropriate. Hence, we focus on a more entangled trial state $|\Psi_\Gamma\rangle$ obtained through the coherent superposition of symmetry projected HF and BdG wavefunctions:

$$|\Psi_\Gamma\rangle = \hat{P}_\Gamma\left(x^{(HF)}\left|\Phi^{(HF)}\right\rangle + x^{(BdG)}\left|\Phi^{(BdG)}\right\rangle\right) \tag{2}$$

Here, $\left|\Phi^{(HF)}\right\rangle = \prod_{n=1}^{N}\hat{c}^\dagger_{\phi_n}|\ \rangle$ with $\hat{c}^\dagger_{\phi_n} = \sum_{\vec{r}\sigma}\hat{c}^\dagger_{\vec{r}\sigma}\phi_{\vec{r}\sigma,n}$ denotes the most general Slater determinant, that mixes both spin components; $\left|\Phi^{(BdG)}\right\rangle \propto \prod_{n=1}^{2N_{\vec{r}}}\hat{\gamma}_n|\ \rangle$ with $\hat{\gamma}_n = \sum_{\vec{r}\sigma}\left(\hat{c}^\dagger_{\vec{r}\sigma}V^*_{\vec{r}\sigma,n} + \hat{c}_{\vec{r}\sigma}U^*_{\vec{r}\sigma,n}\right)$ is the most general quasi-particle vacuum for a lattice with $N_{\vec{r}}$ sites. The Peierls-Yoccoz operator $\hat{P}_\Gamma$ [37] ensures the projection on quantum numbers $\Gamma$ and, according to group theory, may be expressed as a specific linear combination of (unitary) symmetry transformations $\hat{T}_g$:

$$\hat{P}_\Gamma = \sum_g \lambda_{\Gamma,g}\hat{T}_g\,, \tag{3}$$

where the coefficients $\lambda_{\Gamma,g}$ are proportional to the characters of the irreducible representation associated to $\Gamma$. Noting that the transformed vectors $\left|\Phi^{(a)}_g\right\rangle = \hat{T}_g\left|\Phi^{(a)}\right\rangle$ (with the label $a$ specifying the HF or BdG part) remain mean-field states, the variational ansatz Eq. (2) appears as a superposition of numerous symmetry-related wavefunctions. The projected energy $E_\Gamma = \left\langle\hat{H}\right\rangle_{\Psi_\Gamma}$ can also be further calculated according to:

$$E_\Gamma = \frac{\sum_{a,b\in\{HF,BdG\}} x^{(a)*}x^{(b)}\sum_g \lambda_{\Gamma,g}\mathcal{N}^{(a,b)}_g\mathcal{E}\left[\mathcal{R}^{(a,b)}_g\right]}{\sum_{a,b\in\{HF,BdG\}} x^{(a)*}x^{(b)}\sum_g \lambda_{\Gamma,g}\mathcal{N}^{(a,b)}_g} \tag{4}$$

$\mathcal{E}$ stands for the energy functional obtained with Wick's theorem. However, the normal contractions $\left\langle\hat{c}^\dagger_{\vec{r}\sigma}\hat{c}_{\vec{r}'\sigma'}\right\rangle$, $\left\langle\hat{c}_{\vec{r}\sigma}\hat{c}^\dagger_{\vec{r}'\sigma'}\right\rangle$ now correspond to matrix elements between the non-orthogonal wavefunctions $\left|\Phi^{(a)}\right\rangle$ and $\left|\Phi^{(b)}_g\right\rangle$, divided by their overlap $\mathcal{N}^{(a,b)}_g$ [38,39]. They define the one-body (transition) density matrix elements $\left[\rho^{(a,b)}_g\right]_{\vec{r}'\sigma',\vec{r}\sigma}$, $\left[\tilde{\rho}^{(a,b)}_g\right]_{\vec{r}'\sigma',\vec{r}\sigma}$. Similar features apply to the anomalous contributions $\left\langle\hat{c}_{\vec{r}\sigma}\hat{c}_{\vec{r}'\sigma'}\right\rangle$, $\left\langle\hat{c}^\dagger_{\vec{r}\sigma}\hat{c}^\dagger_{\vec{r}'\sigma'}\right\rangle$ that identify to the pairing tensors elements $\left[\kappa_g^{(a,b)}\right]_{\vec r'\sigma',\vec r\sigma}$, $\left[\tilde\kappa_g^{(a,b)}\right]_{\vec r'\sigma',\vec r\sigma}$. Both types of contractions are gathered in the extended matrix $\mathcal{R}_g^{(a,b)} = \begin{pmatrix} \rho_g^{(a,b)} & \kappa_g^{(a,b)} \\ \tilde\kappa_g^{(a,b)} & \tilde\rho_g^{(a,b)} \end{pmatrix}$ that can be easily expressed in terms of quasi-particle states, occupied and unoccupied HF wavefunctions [40]. Stationarity of $E_\Gamma$ Eq. (4) with respect to the amplitudes $x^{(HF)}$ and $x^{(BdG)}$ immediately leads to a generalized eigenvalue equation:

$$\sum_{b\in\{HF,BdG\}} x^{(b)} \sum_g \lambda_{\Gamma,g} \mathcal{N}_g^{(a,b)} \left(\mathcal{E}\left[\mathcal{R}_g^{(a,b)}\right] - E_\Gamma\right) = 0 \tag{5}$$

On the contrary, the energy minimization with respect to the spin-orbitals $\phi_{\vec r\sigma,n}$ and Bogoliubov coefficients $U_{\vec r\sigma,n}$, $V_{\vec r\sigma,n}$ is much more involved and will be detailed in a forthcoming paper [40]. It leads to a set of self-consistent equations that reads:

$$\sum_{b\in\{HF,BdG\}} x^{(b)} \mathcal{L}_\Gamma^{(a,b)} = 0 \tag{6}$$

where the matrices $\mathcal{L}_\Gamma^{(a,b)}$ are obtained with the help of the HF/BdG mean-field Hamiltonian $\mathcal{H}_{ij}[\mathcal{R}] = \frac{1}{2}\frac{\partial \mathcal{E}[\mathcal{R}]}{\partial \mathcal{R}_{ji}}$ as:

$$\mathcal{L}_\Gamma^{(a,b)} = \sum_g \lambda_{\Gamma,g} \mathcal{N}_g^{(a,b)} \left[\left(1-\mathcal{R}_g^{(a,b)}\right)\mathcal{H}\left[\mathcal{R}_g^{(a,b)}\right]\mathcal{R}_g^{(a,b)} + \mathcal{R}_g^{(a,b)}\left(\mathcal{E}\left[\mathcal{R}_g^{(a,b)}\right] - E_\Gamma\right)\right] \tag{7}$$

The system of Eqs. (5)-(7) allows to determine the optimal symmetry projected HF/BdG wavefunction through a numerical solution in which the HF and BdG states are parameterized according to the Thouless theorem [38]. No initial assumption on the ground-state is required and the method is thus able to reveal the physics embedded in the Hubbard model Eq. (1) at low energy.

### 3. Reliability of the HF/BdG approach

We now address the accuracy of the wavefunction Eq. (2) against exact diagonalization (ED) for small clusters or QMC simulations. We focus on autocorrelation functions $\mathcal{M}(\vec r)$, $\mathcal{C}(\vec r)$ and $\mathcal{D}(\vec r)$ in the magnetic, charge and *d*-wave pairing channels, respectively:

$$\mathcal{M}(\vec r) = \left\langle \hat{\vec S}_{\vec 0} \cdot \hat{\vec S}_{\vec r} \right\rangle_{\Psi_\Gamma},\ \mathcal{C}(\vec r) = \left\langle \delta\hat n_{\vec 0} \delta\hat n_{\vec r} \right\rangle_{\Psi_\Gamma},\ \mathcal{D}(\vec r) = \frac{1}{2}\left\langle \hat D_{\vec 0}^\dagger \hat D_{\vec r} + \hat D_{\vec 0} \hat D_{\vec r}^\dagger \right\rangle_{\Psi_\Gamma}. \tag{8}$$

Here, $\hat{\vec S}_{\vec r} = \frac{1}{2}\sum_{\sigma,\sigma'} \hat c_{\vec r\sigma}^+ \vec\tau_{\sigma,\sigma'} \hat c_{\vec r\sigma'}$ is the spin operator at lattice node $\vec r$ (with $\vec\tau$ the usual Pauli matrices); $\delta\hat n_{\vec r} = \sum_\sigma \left(\hat n_{\vec r\sigma} - \langle \hat n_{\vec r\sigma}\rangle_{\Psi_\Gamma}\right)$ corresponds to the local density fluctuation; $\hat D_{\vec r}^\dagger = \sum_{\vec l} f(\vec l) \frac{1}{\sqrt 2}\left(\hat c_{\vec r\uparrow}^\dagger \hat c_{\vec r+\vec l\downarrow}^\dagger - \hat c_{\vec r\downarrow}^\dagger \hat c_{\vec r+\vec l\uparrow}^\dagger\right)$ denotes the singlet pair-field in the $d_{x^2-y^2}$ channel where the form factor $f(\vec l)$ is zero except for neighboring sites in the *x*- and *y*-direction: $f(\pm\vec u_x) = 1$ and $f(\pm\vec u_y) = -1$.

For a $4\times 4$ cluster in the strong coupling regime $U/t = 10,12$, the HF/BdG approximation reproduces very accurately the ED data, as shown in Figs. 1-2. The determination of exact ground-states for larger cells is still limited by the NP-hardness of QMC simulations, except for limited parameter spaces where the stochastic sampling is protected from the notorious sign problem. This is the case at half-filling and we present in Fig. 3 a comparison between QMC and HF/BdG spin-spin correlations for a $6\times 6$ cluster at $U/t=4$. No significant difference is found, especially for the largest separation distances that are essential to indicate the development of a magnetic order. In the hole doped regime with repulsive interactions, a sign-free stochastic sampling of the ground-state is certainly possible [41,42], but it remains generally plagued by systematic errors which origin is not totally elucidated [43]. Nevertheless, it seems that these new QMC algorithms can be accurate for closed-shell fillings and moderate interaction strengths when supplemented by quantum number projection [44,45]. Superfluid correlations in the *d*-wave channel have been investigated in such a framework [45] and we show in Fig. 4 a representative result from Ref. [31] in the intermediate coupling regime $U/t=4$ on a $8\times 8$ cell. The symmetry-adapted HF/BdG wavefunction essentially yields the same pairing response $\mathcal{D}(\vec{r})$, as shown in Fig. 4.

Besides, the variational energies $E_\Gamma$ originating from Eqs. (5)-(7) are summarized in Table 1 for the clusters and on-site interactions previously considered. The agreement is excellent for $4\times 4$ cells with a relative error smaller than 0.5%. The quality of the approximation is quite similar for their doped counterparts, even when a negative next-nearest neighbor hopping $t'$ is introduced to induce frustration. As the size increases, the HF/BdG energy becomes generally less accurate and the deterioration is more pronounced if the cell is doped and/or the coupling $U/t$ is strong. Indeed, while the discrepancy for the half-filled $6\times 6$ cluster at $U/t=4$ does not exceed 0.5%, we only recover an energy $E_\Gamma$ very close to the one obtained with the Gutzwiller projection on top of an optimized BCS wavefunction [28] for 64-sites cells and up to $U/t\sim 12$.

For the largest clusters, the above contradictory findings, regarding the accuracy of HF/BdG correlation functions and energies, could be reconciled provided that improving the ansatz Eq. (2) only has a noticeable effect on the energy. This scenario has been validated by enlarging the variational subspace through the inclusion of several HF/BdG pairs of states via a modification of the wavefunction Eq. (2) according to:

$$\left|\Psi_\Gamma\right\rangle = \hat{P}_\Gamma\left(\sum_{i=1}^{N_{HF/BdG}} \sum_{a\in\{HF,BdG\}} x^{(a_i)}\left|\Phi^{(a_i)}\right\rangle\right) \tag{9}$$

where $\left|\Phi^{(HF_i)}\right\rangle$, $\left|\Phi^{(BdG_i)}\right\rangle$ denote the *i-th* HF, BdG wavefunction in the basis, respectively. The full energy minimization would then require the simultaneous variation of all HF and BdG states. This scheme is beyond our computational facilities and we therefore limit ourselves to a sequential process. In this case, HF/BdG pairs are progressively introduced and each of them is optimized while keeping unchanged the previous basis states. The amplitudes $x^{(a_i)}$ are obtained through a generalized eigenvalue problem similar to Eq. (5) and the set of self-consistent equations determining the structure of the current HF/BdG pair now reads:

$$\sum_{j=1}^{i} \sum_{b\in\{HF,BdG\}} x^{(b_j)}\mathcal{L}_\Gamma^{(a_i,b_j)} = 0 \tag{10}$$

As expected, such superpositions of several HF/BdG wavefunctions notably improve the energy. In the case of a doped $6\times 6$ cluster with $N=24$ atoms and a coupling strength $U/t=8$, the symmetry adapted HF/BdG approximation with one pair of states leads to a variational energy $E_\Gamma=-34.93\,t$. The use of fifty HF/BdG pairs allows to reach an energy of $-36.9\ t$ which is comparable to approximate QMC estimates depending on the constraining state chosen to avoid the sign problem [47]. As shown in Fig. 5, such an improvement of the trial wavefunction Eq. (2) induces minimal changes in the spin, charge and *d*-wave pair correlation functions. A similar behavior is also reported in the next section for larger cells at various fillings.

Manifestly, the ansatz Eq. (9) is not the only way to progressively reconstruct the exact ground-state in a subspace spanned by symmetry projected wavefunctions. As of now, such variational strategies were only developed with HF states, which were either stochastically sampled [48] or optimized [35,36]. The inclusion of BdG wavefunctions in the basis through Eq. (9) yields a notable acceleration of the convergence towards the ground-state. For instance, for $N = 56$ interacting atoms on a $16\times 4$ cell at $U/t = 12$, $E_\Gamma$ is decreased to $-36.018\ t$ with ten HF/BdG pairs while a subspace twice as large is required to reach a similar energy without BdG states [36]. Efficient energy lowering may also be achieved by tuning the numbers of BdG and HF wavefunctions in the basis and the order in which they are introduced as long as the optimization is reduced to a sequential process. As an example, we consider the case of $N = 62$ atoms loaded in a $8\times 8$ cell for $U/t = 8$ through the ansatz

$$\left|\Psi_\Gamma\right\rangle = \hat{P}_\Gamma\left(\sum_{i=1}^{N_{BdG}} x^{(BdG_i)}\left|\Phi^{(BdG_i)}\right\rangle + \sum_{i=1}^{N_{HF}} x^{(HF_i)}\left|\Phi^{(HF_i)}\right\rangle\right). \tag{11}$$

While the simple HF/BdG state Eq. (2) gives an energy $E_\Gamma = -34.736\ t$ not competitive with extended BCS-Gutzwiller schemes [49], the expansion Eq. (11) allows to reach a similar accuracy with $E_\Gamma = -35.961\ t$ (for $N_{BdG} = 15$ and $N_{HF} = 35$ ). Fig. 6 reconfirms that the physical content embedded in $\mathcal{M}(\vec{r})$, $\mathcal{C}(\vec{r})$ and $\mathcal{D}(\vec{r})$ is unaffected against the enlargement of the HF/BdG subspace. However, noticeable changes in the values of the order parameters extracted from the long-ranged parts of the considered correlation functions are obtained.

Finally, the present calculations tend to support the HF/BdG approximation with full symmetry restoration before variation as a reliable starting point to capture the essence of correlations entailed in the repulsive Hubbard model, at least in the magnetic, density, and superfluid channels and for moderate size clusters.

## 4. Results: Quantum phase diagram of ultracold fermions loaded in optical four-leg tubes.

Though the symmetry projected HF/BdG wavefunction Eq. (2) displays a polynomial complexity with the number $N_{\vec{r}}$ of lattice sites, the numerical optimization remains challenging by requiring the simultaneous determination of around $3N_{\vec{r}}^2$ parameters. Unfortunately, very large square cells are needed to support both the emergence of an off-diagonal long-ranged order linked with superfluidity and the development of long wavelength collective modes expected in the density and magnetic channels [32]. Therefore, we now restrict ourselves to four-leg ladders that are natural steps in the dimensional crossover from the exactly solvable chain to the unknown 2D limit. This geometry can be in fact indirectly emulated with ultracold vapors by loading the atoms in optical tubes of plaquettes created from four wells arranged in a square pattern, as depicted in Fig. 7. A three-dimensional array of such independent clusters has already been realized using an optical superlattice configuration along two orthogonal directions [24]. By tuning the laser potentials parameters to allow for the tunneling between adjacent planes, a collection of uncoupled identical tubes could be obtained. When unfolded, each of them realizes the Hubbard Hamiltonian on a rectangular cell with four legs and periodic boundary conditions along the $y$-direction.

The variational state Eq. (2), free of symmetry breaking, is now systematically determined to unravel the relevant orders and their potential intertwining in the low-lying energy states. We focus on tubes of length $L \geq 16$ loaded with slightly less than one atom per site, so that they are characterized by their hole doping $\delta = 1 - n$ with $n$ the lattice filling factor. Coupling strengths ranging from the moderate $(U/t = 4)$ to the strongly $(U/t = 12)$ interacting regime are considered. We stress that all energy minimizations have been independently carried out, thereby allowing for crosschecking the results. Moreover, full periodic boundary conditions on finite-size clusters could bias pairing correlations by disadvantaging the $d_{x^2-y^2}$ channel: The corresponding wavefunction in momentum space would indeed be zero for a non-negligible fraction of wavevectors in the first Brillouin zone. Therefore, antiperiodic boundary conditions along the legs ($x$-direction) are chosen. Their influence is discussed in the Appendix, where it is more generally shown that a tube length $L = 16$ is large enough to ensure a weak sensitivity of the physical content to boundary conditions.

We first address a system of $L = 16$ four-sites plaquettes with an on-site interaction $U = 12t$ and investigate the hole doping dependence of relevant correlation functions. Each optimization involves the determination of around $10^4$ complex parameters that enter the variational wavefunction Eq. (2), consisting of the coherent superposition of more than $7\ 10^4$ symmetry related mean-field states. The resulting magnetic $S_m(\vec{q})$ and density $S_c(\vec{q})$ structure factors are shown in Fig. 8. They are defined according to:

$$S_m(\vec{q}) = \frac{4}{3}\sum_{\vec{r}} \exp(i\vec{q}\cdot\vec{r})\ \mathcal{M}(\vec{r}),\ S_c(\vec{q}) = \sum_{\vec{r}} \exp(i\vec{q}\cdot\vec{r})\ \mathcal{C}(\vec{r}). \tag{12}$$

Pairing correlation functions $\mathcal{D}(\vec{r})$ in the $d$-wave channel are displayed in Fig. 9 [50].

Starting with $\delta \approx 16\%$ $(N = 54)$, a coexistence of spin and charge density waves is clearly evidenced by a peak in $S_m$ and $S_c$ on top of a broad background. The dominant wavevector $\vec{q}_m = (3\pi/4, \pi)$ in the spin-spin correlations corresponds to an antiferromagnet with a staggered magnetization oscillating in amplitude with a period of $\lambda_m = 8$ lattice spacings in the $x$-direction. Similarly, the density-density correlation function reveals inhomogeneities distributed with a period $\lambda_c = 4$ along the $x$-axis in the variational ground-state. Note that these orders and their symmetry-related counterparts are necessarily superimposed to respect all invariances of the Hamiltonian. Furthermore, with $\lambda_c$ even, the relation $\lambda_m = 2\lambda_c$ characterizes stripes at the boundaries of antiferromagnetic domains separated by a $\pi$ phase shift. Their intertwining with $d$-wave superfluidity is eventually proved by highlighting in Fig. 9b a non-zero average of the pairing correlation function $\mathcal{D}(\vec{r})$ at large separation distance $r$. The non-decaying tail observed for $r > 4$ is consistent with off-diagonal long-ranged order that signs superfluidity. Besides, the 4-period small oscillations of $\mathcal{D}(\vec{r})$ around its averaged value indicate the existence of pairs at a finite momentum equal to the charge-order wavevector. Such stripes with a $d$-wave superfluidity spatially modulated in phase with the density profile have also been proposed in recent simulations [51,52] of the $t$-$J$ Hamiltonian that approximates the Hubbard model in the limit $U/t \to \infty$. Superfluid domain wall states in four-leg ladders also find support from density-matrix renormalization group calculations [53] of the *t-J* model, despite of a possible contamination by Friedel oscillations stemming from open boundary conditions [54].

Stripe-like states are robust against a decrease of the hole number as shown in Fig. 8 for the $16\times 4$ cluster considered here. However, the shift of the peaks in the structure factors $S_m$ and $S_c$ reflects a doubling of the period when crossing $\delta = 1/8$. In addition, pairing correlations at large distance are totally suppressed in the 8 and 4 hole systems corresponding to perfectly filled and half-filled 8-period vertical stripes, respectively (see Fig. 9c). The spin and charge pattern associated to domain walls separated by eight lattice spacings is also realized for $N = 58$. As for $N = 54$, these stripes are neither filled nor half-filled and again the behavior of $\mathcal{D}(\vec{r})$ at large distance is consistent with the development of a pair-density wave of period $\lambda_c$. When moving towards the half-filling limit, antiferromagnetism no longer exhibits amplitude modulation and a uniform density profile is recovered. Finally, a pure *d*-wave off-diagonal long-ranged order is unambiguously supported as long as such a background is doped with few holes (see Fig. 9b for $N = 62$).

Another scenario emerges when considering an increase of the hole doping from $\delta \approx 16\%$. While peaks related to charge-density waves disappear, incommensurate spin-spin correlations persist. At the same time, its associated wavevector leaves the side of the Brillouin zone to its diagonal. The nature of the underlying incommensurate magnetic ordering is not unambiguously revealed by such peaks, as they are compatible with both collinear spins or spirals [29]. One way to test whether spins rotate on the lattice is to detect a non-decaying four-body correlation function between spin chirality vectors $\hat{\vec{V}}_{\vec{r}} = \hat{\vec{S}}_{\vec{r}} \wedge \left( \hat{\vec{S}}_{\vec{r}+\vec{u}_x} + \hat{\vec{S}}_{\vec{r}+\vec{u}_y} \right)$ as a function of separation distance. The calculation of $\mathcal{V}(\vec{r}) = \left\langle \hat{\vec{V}}_{\vec{0}} \cdot \hat{\vec{V}}_{\vec{r}} \right\rangle_{\Psi_\Gamma}$ at $U = 12t$ for different densities is shown in Fig. 10. The long-ranged $(r > 4)$ part systematically displays an oscillating behavior reflecting significant quantum fluctuations. Two regimes are however clearly distinguished: Spiral correlations averaged over large distances vanish in striped and antiferromagnetic states $(N \geq 54)$, while they are non-zero and positive at larger dopings. This signal remains of small amplitude and thus rather characterizes a spiral ordering component embedded in a spin-density wave (SDW). Note that anyhow, pure spiral ground-states are not expected in the large-$U$ Hubbard model considered here [55]. As shown in Fig. 9a, the *d*-wave pairing correlation function in the SDW/Spiral state displays a complex behavior at large distance, yet free of a rapid decay to zero as was found at half-filling or in the stripes at commensurate dopings. It can be viewed as the precursor of the *d*-wave superfluidity that is better established for larger lattice fillings.

The energy minimization with the symmetry projected HF/BdG wavefunction essentially exhibits all the above features from the intermediate coupling $U/t = 6$ to the strongly correlated regime $U/t = 12$. The results are summarized in the quantum phase diagram shown in Fig. 11 for hole doping $\delta$ smaller than $1/4$. Stripe-like states are stabilized in the intermediate doping range and once $U/t$ exceeds a critical value. The latter is suppressed with decreasing $\delta$. A similar feature has also been obtained with inhomogeneous dynamical mean-field [56] and constrained-path QMC approaches [32]. In addition, the change of the charge period from $\lambda_c = 4$ to $\lambda_c = 8$ takes place for $U \geq 10t$ when crossing $\delta = 1/8$. Close to the half-filling limit, only antiferromagnetic correlations persist, while stripes melt for larger doping. Instead, incommensurate antiferromagnetism in the form of coexisting spiral and spin-density waves is found. It develops along the *x*-direction for intermediate interaction strengths $(6 \le U/t \le 8)$ and tends towards the diagonal direction at large $U/t$. Furthermore, the spiral component appears for couplings that increase with the doping. Finally, long-ranged *d*-wave pairing correlations are systematically evidenced, except when all the holes are perfectly trapped into filled or half-filled vertical stripes. These trends are altered at smaller $U/t$. In particular, for $U/t = 4$, charge inhomogeneities are missing and a clear tendency towards magnetic ordering is obtained for doping $\delta < 16\%$, only, in agreement with latest diagrammatic QMC calculations [20]. Eventually, the superfluid signal is rather erratic, though this non-monotonicity proved stable against changes of boundary conditions to investigate the influence of shell effects, commonly invoked at small coupling in the attractive regime [57]. Further details are presented in the Appendix.

The symmetry projected HF/BdG phase diagram Fig. 11 in four-leg ladder geometry confirms the emergence of correlations proposed separately for the hole-doped 2D Hubbard model in the spin, charge and pairing channels. While the scenario of a competition between the resulting orders is usually retained, our findings rather point towards a subtle entanglement of the associated degrees of freedom. It induces the wide variety of strongly correlated states observed in Fig. 11 as a function of the hole doping. Their robustness requires persistence of the observed correlations when refining the grid of available densities by increasing the tube length $L$. Some representative examples are shown in Fig. 12 and Fig. 13 for different $U/t$ regimes to explore additional parts of the phase diagram originally obtained at $L = 16$. Either close to the half-filling point or on both sides of the $1/8$ hole doping, no qualitatively new features appear in the spin and density autocorrelation functions. Not even is the stripe period changed, when relevant. Besides, long-ranged pair correlations are still evidenced whether they are intertwined with antiferromagnetism or stripes. In the latter case, the increase of $L$ allows to grasp the oscillations of $\mathcal{D}(\vec{r})$ which clearly match the charge period $\lambda_c$. It is remarkable that only the quantum number projection on top of mean-field like wavefunctions and before variation remains efficient to generate such unconventional superfluid signal for clusters with a hundred of lattice sites. Indeed, we recall that the standard BdG approach fails to stabilize superfluid states for the repulsive Hubbard model. However, the value of the long-ranged tail in $\mathcal{D}(\vec{r})$ tends to decrease as compared to the length $L = 16$ previously considered. While this feature could indicate the establishment of a quasi-long-ranged order in the *d*-wave pairing channel, the benchmark comparisons presented in Section 3 point towards a deterioration of the symmetry projected HF/BdG approximation to the ground-state when enlarging the cluster. So, one cannot exclude the need to significantly increase the dimension of the HF/BdG subspace to recover the accuracy on correlation functions reached for $L = 16$. Such calculations are not currently feasible and would then reveal a reminiscence of the intrinsic exponential complexity met by unbiased methods.

## 5. Conclusion

Summarizing, we have highlighted insights into generic features of repulsively interacting ultracold fermions loaded in optical four-leg ladders through their description by the Hubbard model. First, we have shown that such systems are ideal candidates to realize a whole sequence of magnetic phases that may be tailored by varying the filling of the lattice or the ratio $t/U$. Above all, the since long proposed scenario of *d*-wave superfluidity emerging from a doped Mott insulator has been put forward thanks to energy minimizations with no physical assumption on the relevant orders. Nevertheless, such intertwining of magnetic and pair degrees of freedom manifests itself under various facets depending on whether antiferromagnetic correlations grow from homogenous collinear spins, spatially modulated spin-density waves or spirals. It also involves the charge degree of freedom as stripes that either destroy or support superfluidity, depending on their filling. These features have been extracted from symmetry-adapted states originating from quantum number projection that also induce correlations beyond mean-field. Furthermore, magnetic, charge and superfluid correlations remain robust against improvements of this wavefunction. The quantum phase diagram in the four-leg tube geometry therefore provides an additional reference for the cross-validation between theory and quantum emulation from experiments that is necessary to face the exponential complexity of low dimensional quantum matter.

**Acknowledgments**

We warmly thank T. Kopp and D. Braak for stimulating discussions. This research was partly supported by the ANR through the GeCoDo project (ANR-11-JS08-001-01). We are grateful to the Région Basse-Normandie and the Ministère de la Recherche for financial support.

**Appendix**

We discuss here the influence of the boundary conditions along the tube direction in the symmetry projected HF/BdG approximation to the ground-state of the Hubbard model in a four-leg geometry. Emphasis is put on the energy $E_\Gamma$, the magnetic $S_m(\vec{q})$ and charge $S_c(\vec{q})$ structure factors Eqs. (12), (8) as on the correlation function $\mathcal{D}(\vec{r})$ in the *d*-wave pairing sector Eq. (8). The results obtained with periodic (PBC) and antiperiodic (APBC) boundary conditions are compared through several representative cases for a tube length $L = 16$.

We first focus on two neighboring fillings $N = 56$ and $N = 58$ in the intermediate regime $U/t = 4$ to probe the robustness of the non-monotonicity of the *d*-wave superfluid response reported in the phase diagram Fig. 11 for this interaction. Indeed, the magnetic and density correlations reported on Fig. 14 reveal only marginal differences between PBC and APBC. In both cases, a spin-density wave with the magnetic period $\lambda_m = 16$ is found. Regarding the pairing correlations in the *d*-wave channel, no sensitivity to the tube boundaries appears when $\mathcal{D}(\vec{r})$ is essentially zero at large distance, as shown in Fig. 14 for $N = 56$. On the contrary, precursors of *d*-wave superfluid states are more subject to be influenced by the choice of PBC or APBC, as anticipated in Sec. 4. This is clearly the case for $N = 58$ where PBC maintains a long-ranged plateau in $\mathcal{D}(\vec{r})$, but with a reduced value. When moving to the strongly correlated regime, the superfluid behavior displays a similar effect against the change of boundary conditions as long as it is intertwined with long wavelength modes in the magnetic and/or density channels. For instance, in the paired-stripe state obtained for $N = 58$ atoms at $U/t = 10$ and shown in Fig. 15a, $\mathcal{D}(\vec{r})$ with PBC or APBC exhibits an oscillating shape at large $r$ though PBC significantly reduces the amplitude as well as the averaged value. On the other hand, when considering $N = 62$ atoms that realize a lightly hole doped antiferromagnetic Mott insulator, $\mathcal{D}(\vec{r})$ is no longer affected whether PBC or APBC are selected (see Fig 15b).

Finally, the present symmetry projected HF/BdG calculations suggest that the main results summarized in the phase diagram Fig. 11 are not significantly contaminated from boundary condition effects.

[1] Fradkin E and Kivelson S A 2012 *Nature Phys*. **8** 865
[2] Bloch I, Dalibard J and Zwerger W 2008 *Rev. Mod. Phys.* **80** 885
[3] Zwierlein M W, Abo-Shaeer J R, Schirotzek A, Schunck C H and Ketterle W 2005 *Nature* **435** 1047
[4] Zwierlein M W, Schunck C H, Schirotzek A and Ketterle W 2006 *Nature* **442** 54
[5] Sanner C, Su E J, Huang W, Keshet A, Gillen J and Ketterle W 2012 *Phys. Rev. Lett.* **108** 240404
[6] Jo G B, Lee Y R, Choi J H, Christensen C A, Kim T H, Thywissen J H, Pritchard D E and Ketterle W 2009 *Science* **325** 1521
[7] Lewenstein M, Sampera A, Ahufinger V, Damski B, Sen A and Sen U 2007 *Adv. Phys.* **56** 243
[8] Köhl M, Moritz H, Stöferle T, Günter K and Esslinger T 2005 *Phys. Rev. Lett.* **94** 080403
[9] Chin J K, Miller D E, Liu Y, Stan C, Setiawan W, Sanner C, Xu K and Ketterle W 2006 *Nature* **443** 961
[10] Hubbard J 1963 *Proc. Roy. Soc. London A* **276** 238
[11] Loh Y L and Trivedi N 2010 *Phys. Rev. Lett.* **104** 165302
[12] Sewer A, Zotos X and Beck H 2002 *Phys. Rev. B* **66** 14504(R)
[13] Anderson P W 1987 *Science* **235**, 1196
[14] Xu Z, Stock C, Chi S, Kolesnikov A I, Xu, G, Gu G and Tranquada J M 2014 *Phys. Rev. Lett.* **113** 177002
[15] Fradkin E, Kivelson S A and Tranquada J M 2015 *Rev. Mod. Phys.* **87** 457
[16] Lieb E H and Wu F Y 1968 *Phys. Rev. Lett.* **20** 1445
[17] Bulla R, Costi T and Vollhardt D 2001 *Phys. Rev. B* **64** 045103
[18] Troyer M and Wiese U J 2005 *Phys. Rev. Lett*. **94** 170201
[19] Liang S and Pang H 1994 Phys. Rev. B **49** 9214
[20] Deng Y, Kozik E, Prokof'ev N and Svistunov B 2015 *EPL* **110** 57001
[21] Jördens R, Strohmaier N, Günter K, Moritz H and Esslinger T 2008 *Nature* **455**, 204
[22] Hart R A, Duarte P M, Yang T L, Liu X, Paiva T, Khatami E, Scalettar R T, Trivedi N, Huse D A and Hulet R G 2015 *Nature* **519** 211
[23] Van Houcke K, Werner F, Kozik E, Prokof'ev N, Svistunov B, Ku M J H, Sommer A T, Cheuk L W, Schirotzek A, Zierlein M 2012 *Nature Phys.* **8** 366
[24] Nascimbène S, Chen Y A, Atala M, Aidelsburger M, Trotzky S, Paredes B and Bloch I 2012 *Phys. Rev. Lett.* **108** 205301
[25] Halboth C J and Metzner W 2000 *Phys. Rev. B* **61** 7364
[26] Metzner W, Salmhofer M, Honerkamp C, Meden V and Schönhammer J 2012 *Rev. Mod. Phys*. **84** 299
[27] Vollhardt D 1984 *Rev. Mod. Phys.* **56** 99
[28] Giamarchi T and Lhuillier C 1991 *Phys. Rev. B* **43** 12943
[29] Dzierzawa M and Frésard R 1993 *Z. Phys. B* **91** 245
[30] Miyazaki M, Yamaji K, Yanagisawa T and Kadono R 2009 *J. Phys. Soc. Jpn* **78** 043706
[31] Misawa T and Imada M 2014 *Phys. Rev. B* **90** 115137
[32] Chang C C and Zhang S 2010 *Phys. Rev. Lett.* **104** 116402
[33] Juillet O and Frésard R 2013 *Phys. Rev B* **87** 115136
[34] Leprévost A, Juillet O and Frésard R 2014 *Ann. Phys. (Berlin)* **526** 430
[35] Tomita N and Watanabe S 2009 *Phys. Rev. Lett.* **103** 166401
[36] Rodríguez-Guzmán R, Jiménez-Hoyos C and Scuseria G E 2014 *Phys. Rev. B* **90** 195110
[37] Hamermesh M 1962 *Group Theory and its Applications to Physical Problems* (Reading MA: Addison-Wesley)

[38] Blaizot J P and Ripka G 1986 *Quantum Theory of Finite Systems* (Cambridge, MA: MIT Press)

[39] The overlap $\mathcal{N}_g^{(a,b)} = \left\langle \Phi^{(a)} \middle| \Phi_g^{(b)} \right\rangle$ is a determinant for two HF wavefunctions (see Ref. [38]) and a pfaffian otherwise (see Bertsch G F and Robledo L M 2012 *Phys. Rev. Lett.* **108** 042505).

[40] Leprévost A, Juillet O and Frésard F (2016) *in preparation*

[41] Corney J F and Drummond P D 2004 *Phys. Rev. Lett* **93** 260401

[42] Juillet O 2007 *New J. Phys*. **9** 163

[43] Assaad F F, Werner P, Corboz P, Gull E and Troyer M 2005 *Phys. Rev. B* **72** 22451

[44] Aimi T and Imada M 2007 *J. Phys. Soc. Jpn.* **76** 084709

[45] Aimi T and Imada M 2007 J. Phys. Soc. Jpn. **76** 113708

[46] Bauer B, Carr L D, Evertz H G, Feiguin A, Freire J, Fuchs S, Gamper L, Gukelberger J, Gull E, Guertler S, Hehn A, Igarashi R, Isakov S V, Koop D, Ma, P N, Mates P, Matsuo H, Parcollet O, Pawlowski G, Picon J D, Pollet L, Santos E, Scarola V W, Schollwöck U, Silva C, Surer B, Todo, S, Trebst S, Troyer M, Wall M L, Werner P and Wessel S 2011 *J. Stat. Mech.* **P05001**

[47] Shi H, Jimenez-Hoyos C A, Rodriguez-Guzman R, Scuseria G E and Zhang S 2014 *Phys. Rev. B* **89** 125129

[48] Misuzaki T and Imada M 2004 *Phys. Rev. B* **69** 125110

[49] Eichenberger D and Baeriswyl D 2007 *Phys. Rev. B* **76** 180504

[50] The rapidly decaying one-particle contributions in the pairing correlation function $\mathcal{D}(\vec{r})$ are discarded to avoid any spurious contamination from non-interacting dressed atoms. Precisely, the quantity $\left\langle \hat{c}^\dagger_{\vec{r}_1\sigma_1} \hat{c}_{\vec{r}_4\sigma_4} \right\rangle_{\Psi_\Gamma} \left\langle \hat{c}^\dagger_{\vec{r}_2\sigma_2} \hat{c}_{\vec{r}_3\sigma_3} \right\rangle_{\Psi_\Gamma} - \left\langle \hat{c}^\dagger_{\vec{r}_1\sigma_1} \hat{c}_{\vec{r}_3\sigma_3} \right\rangle_{\Psi_\Gamma} \left\langle \hat{c}^\dagger_{\vec{r}_2\sigma_2} \hat{c}_{\vec{r}_4\sigma_4} \right\rangle_{\Psi_\Gamma}$ is subtracted for each term of the form $\left\langle \hat{c}^\dagger_{\vec{r}_1\sigma_1} \hat{c}^\dagger_{\vec{r}_2\sigma_2} \hat{c}_{\vec{r}_3\sigma_3} \hat{c}_{\vec{r}_4\sigma_4} \right\rangle_{\Psi_\Gamma}$, so that $\mathcal{D}(\vec{r})$ vanishes in the $U = 0$ limit.

[51] Corboz P, White S R, Vidal G and Troyer M 2011 *Phys. Rev. B* **84** 041108

[52] Corboz P, Rice T M and Troyer M 2014 *Phys. Rev. Lett.* **113** 046402

[53] White S R and Scalapino D J 1997 Phys. Rev. B **55** R14701

[54] White S R, Affleck I and Scalapino D J 1997 Phys. Rev. B **65** 165122

[55] Raczkowski M, Frésard R and Oles A M 2006 *Europhys. Lett.* **76** 128

[56] Peters R and Kawakami N 2014 *Phys. Rev. B* **89** 155134

[57] Bormann D, Schneider T and Frick M 1991 *Europhys. Lett.* **14** 101

**Table 1**. Variational energies $E_\Gamma$ from the symmetry projected HF/BdG wavefunction compared to reference energies $E_{ref.}$ obtained either with exact diagonalization (ED), quantum Monte-Carlo (QMC) or variational Monte-Carlo (VMC) results. Periodic-periodic (PP) or periodic-antiperiodic (PA) boundary conditions are specified. The symbol (*) indicates a simulation of the frustrated Hubbard model with an hopping amplitude $t' = -0.3t$ between next-nearest neighbors. Exact diagonalization has been performed with ALPS [46]. QMC data are borrowed from Ref. [31,43]. The VMC calculations correspond to the original (o) [28] or improved (i) [31,49] BCS-Gutzwiller wavefunction.

| Lattice | $U/t$ | Boundary | $N$ | $E_\Gamma/t$ | $E_{ref.}/t$ |
|---|---|---|---|---|---|
| $4\times4$ | 4 | PP | 16 | −13.618 | −13.622 (ED) |
| $4\times4$ (*) | 8 | PP | 14 | −12.439 | −12.503 (ED) |
| $4\times4$ | 10 | PP | 10 | −16.876 | −16.902 (ED) |
| $4\times4$ | 12 | PP | 14 | −9.957 | −10.05 (ED) |
| $6\times6$ | 4 | PP | 36 | −30.724 | −30.87(2) (QMC) |
| $8\times8$ | 4 | PP | 50 | −70.13 | −71.417(4) (VMC, i)<br>−72.51(5) (QMC) |
| $8\times8$ | 10 | PA | 60 | −32.164 | −31.2 (VMC, o) |
| $8\times8$ | 8 | PA | 62 | −34.736 | −36.04 (VMC, i) |

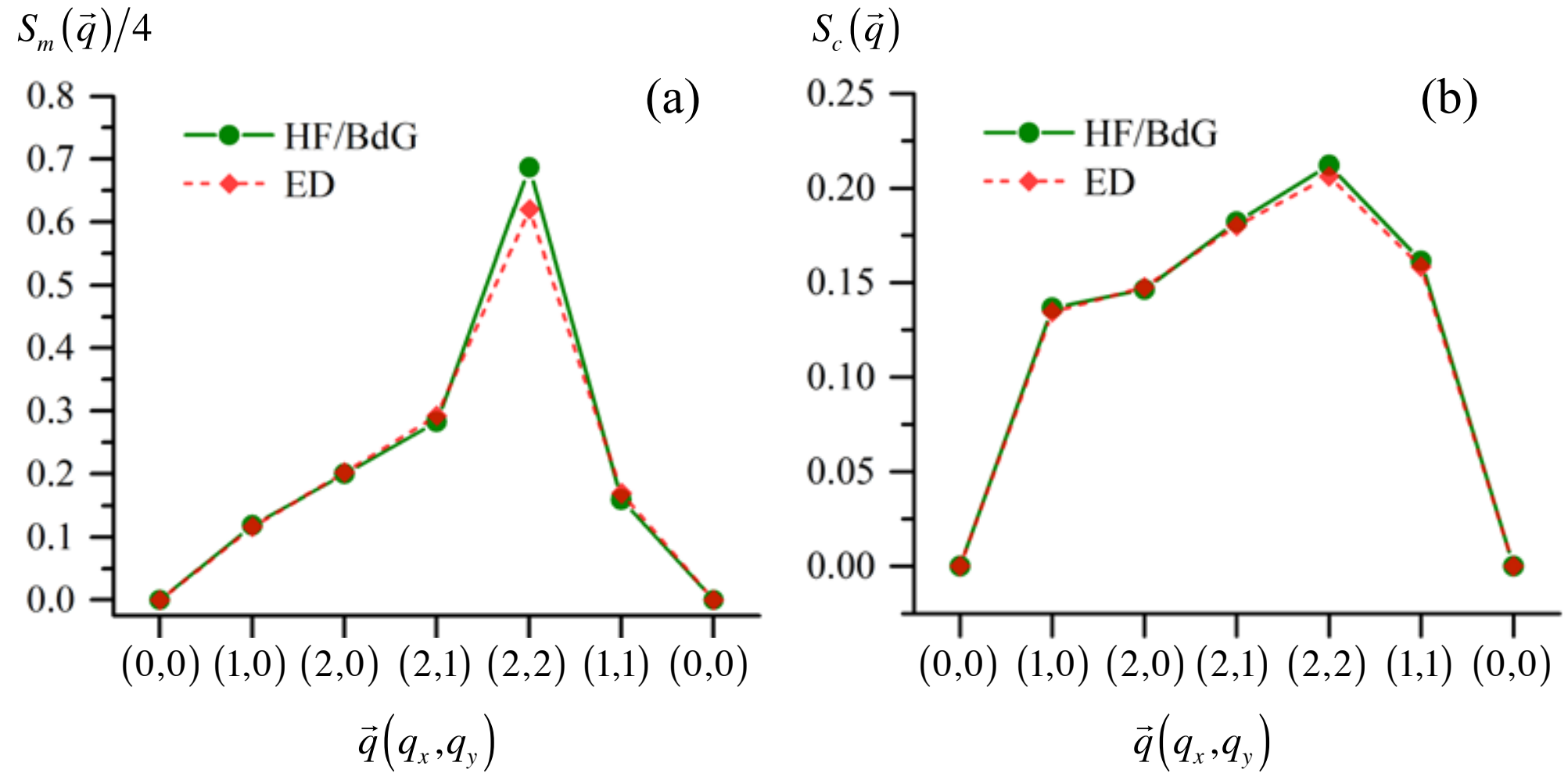

**Figure 1.** Momentum dependence of (a) magnetic $S_m(\vec{q})$ and (b) charge $S_c(\vec{q})$ structure factors for a $4\times4$ cluster with $N=14$ atoms and periodic-periodic boundary conditions in the strongly correlated regime $U/t=12$. Wavevectors $\vec{q}$ are expressed in units of $\pi/2$. A symmetry projected HF/BdG pair of states displays excellent agreement with exact diagonalization.

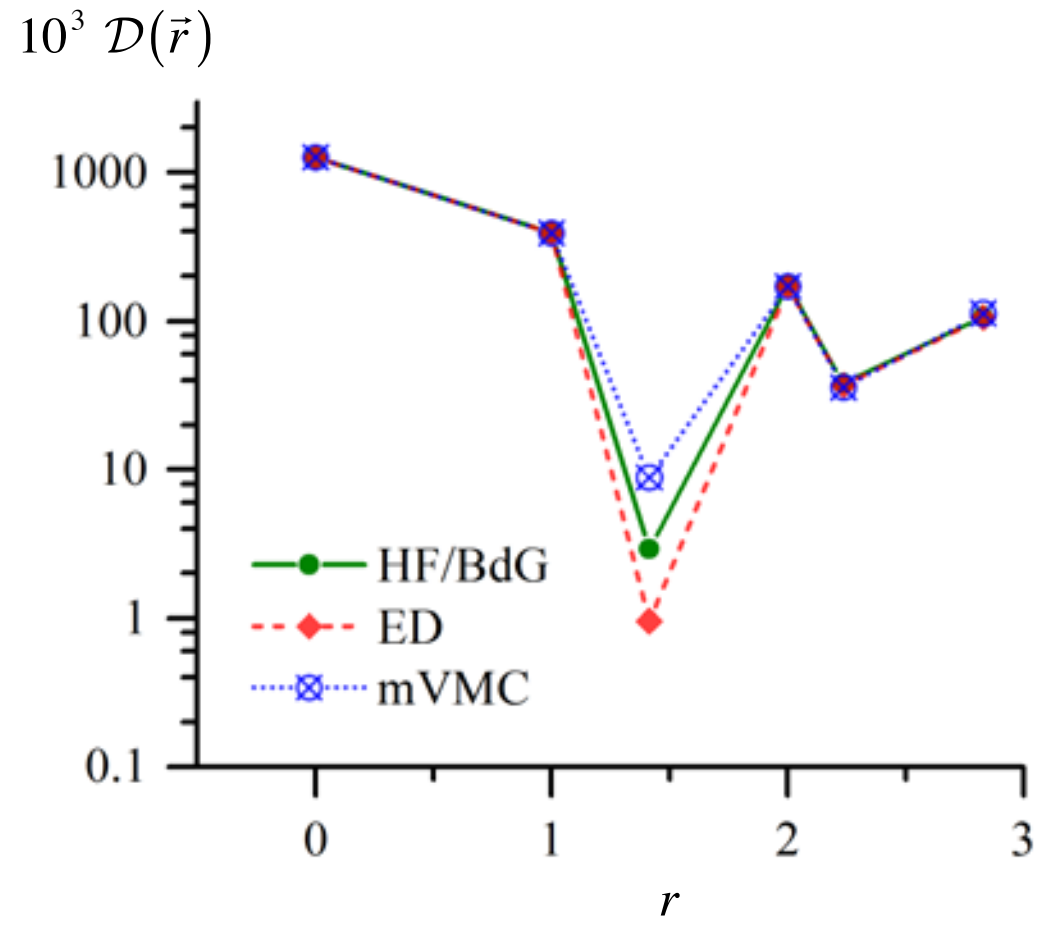

**Figure 2.** $d$-wave pair correlation function $\mathcal{D}(\vec{r})$ against separation distance $r$ for a $4\times4$ lattice with $N=10$ atoms at strong on-site interaction $U/t=10$. Periodic-periodic boundary conditions are imposed. ED results as well as a recent VMC calculation with a symmetry restored BCS-Gutzwiller wavefunction (mVMC) are extracted from Fig. 8b of Ref. [31]. The HF/BdG ansatz, with quantum number projection before variation, correctly reproduces the shape and magnitude of the exact pairing correlations.

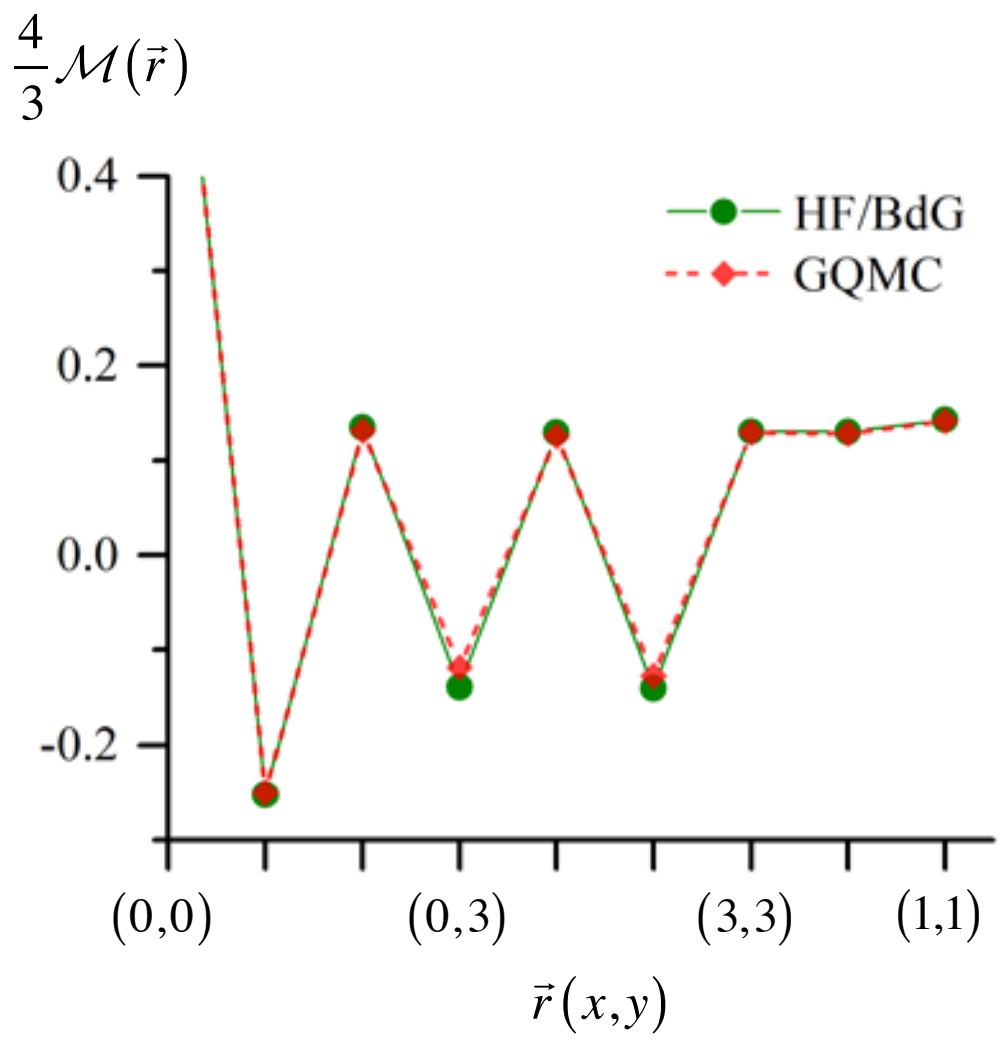

**Figure 3.** Spin-spin correlations $\mathcal{M}(\vec{r})$ at half-filling for a periodic-periodic $6\times 6$ cluster at $U/t = 4$ as obtained from the symmetry restored HF/BdG approach and compared with QMC calculations incorporating quantum number projection (GQMC, extracted from Fig. 2 of Ref. [43]).

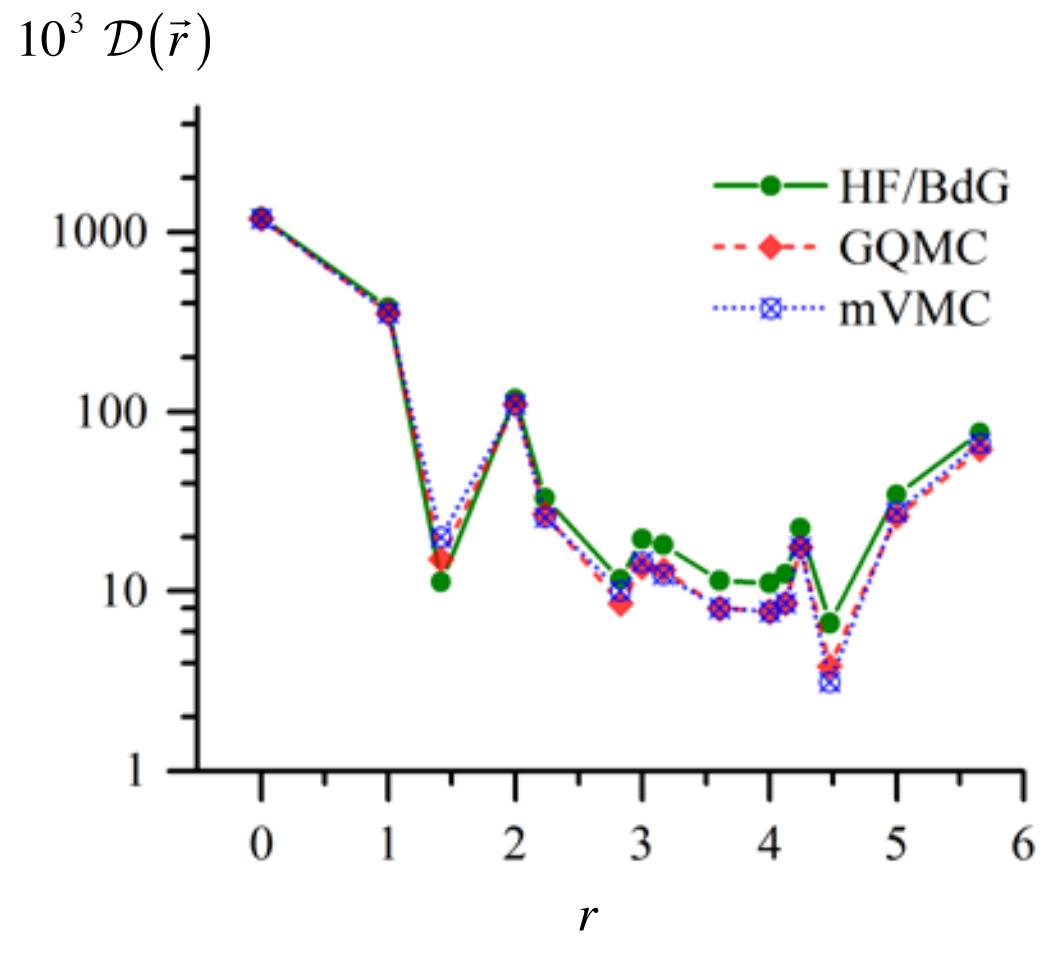

**Figure 4.** Distance dependence of the *d*-wave pair correlation function $\mathcal{D}(\vec{r})$ for a $8\times 8$ cell with $N = 50$ atoms and periodic-periodic boundary conditions at moderate coupling strength $U/t = 4$. Sign-free QMC calculations (GQMC) and VMC results with the BCS-Gutzwiller wavefunction (mVMC), both including symmetry restoration, are extracted from Fig. 10 of Ref. [31]. They are compared to the pairing correlations $\mathcal{D}(\vec{r})$ originating from the symmetry adapted HF/BdG scheme.

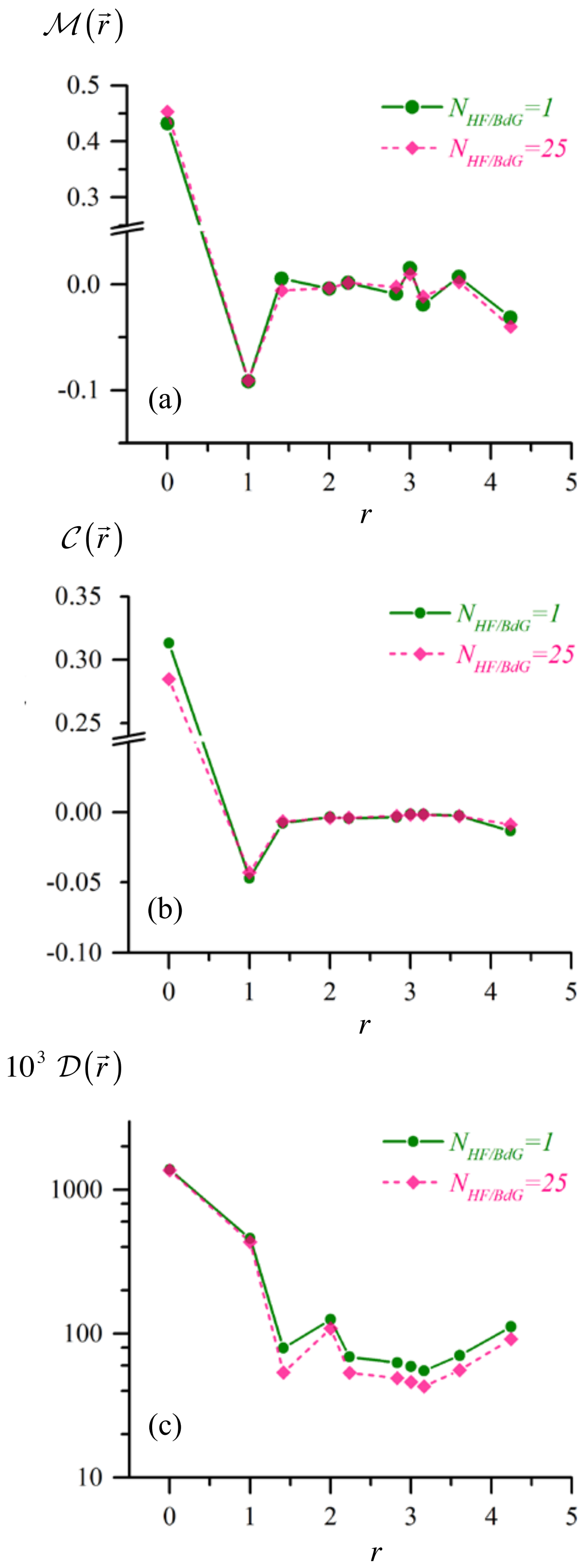

**Figure 5.** Evolution of (a) spin, (b) density and (c) *d*-wave pairing autocorrelation functions with the number $N_{HF/BdG}$ of symmetry projected HF/BdG wavefunctions spanning the variational subspace. Calculations are performed for $N = 24$ atoms on a $6\times 6$ cluster with an interaction strength $U/t = 8$.

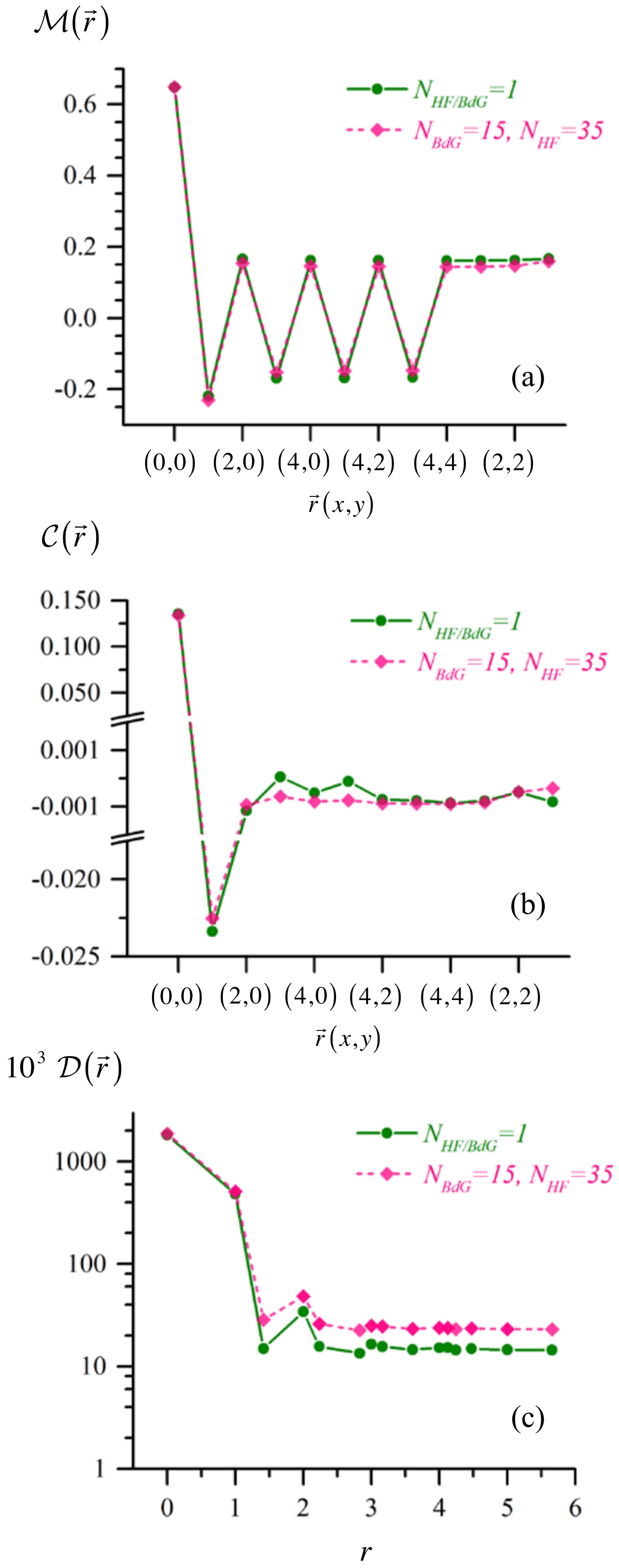

**Figure 6.** Spatial dependence of (a) spin-spin, (b) density-density and (c) *d*-wave pair-pair correlations obtained from the two different symmetry projected wavefunctions Eq. (2) and Eq. (11). $N=62$ interacting atoms in the regime $U/t=8$, and loaded on a $8\times 8$ cell are considered.

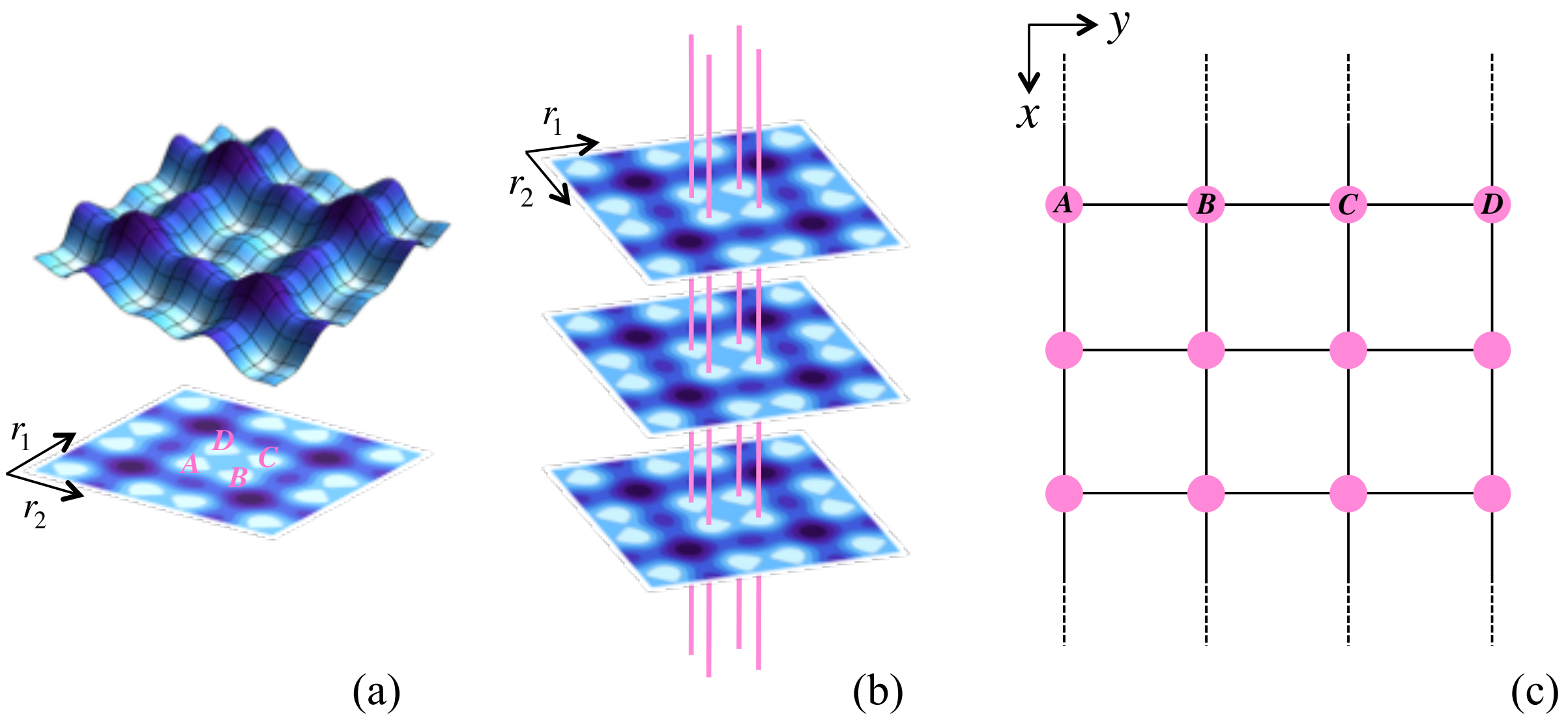

**Figure 7.** Representation of an optical superlattice configuration emulating the four-leg tube Hubbard model. (a) Scheme of the potential $V(r_1,r_2)=-\sum_{i=1}^{2}\left(V_l\sin^2(kr_i)+V_s\sin^2(2kr_i)\right)$ generated in a horizontal $r_1-r_2$ plane by two mutually orthogonal pairs of standing-wave laser fields, each with a wavelength ratio of 2. With this setup, a four-site unit cell $ABCD$ is created. (b) By superimposing another optical lattice along the vertical direction, independent tubes of plaquettes may be obtained. (c) When using periodic boundary conditions along the $y$-direction in a $x-y$ lattice, the $L\times 4$ rectangular cluster is isomorphic to each optical four-leg ladder displayed in (b).

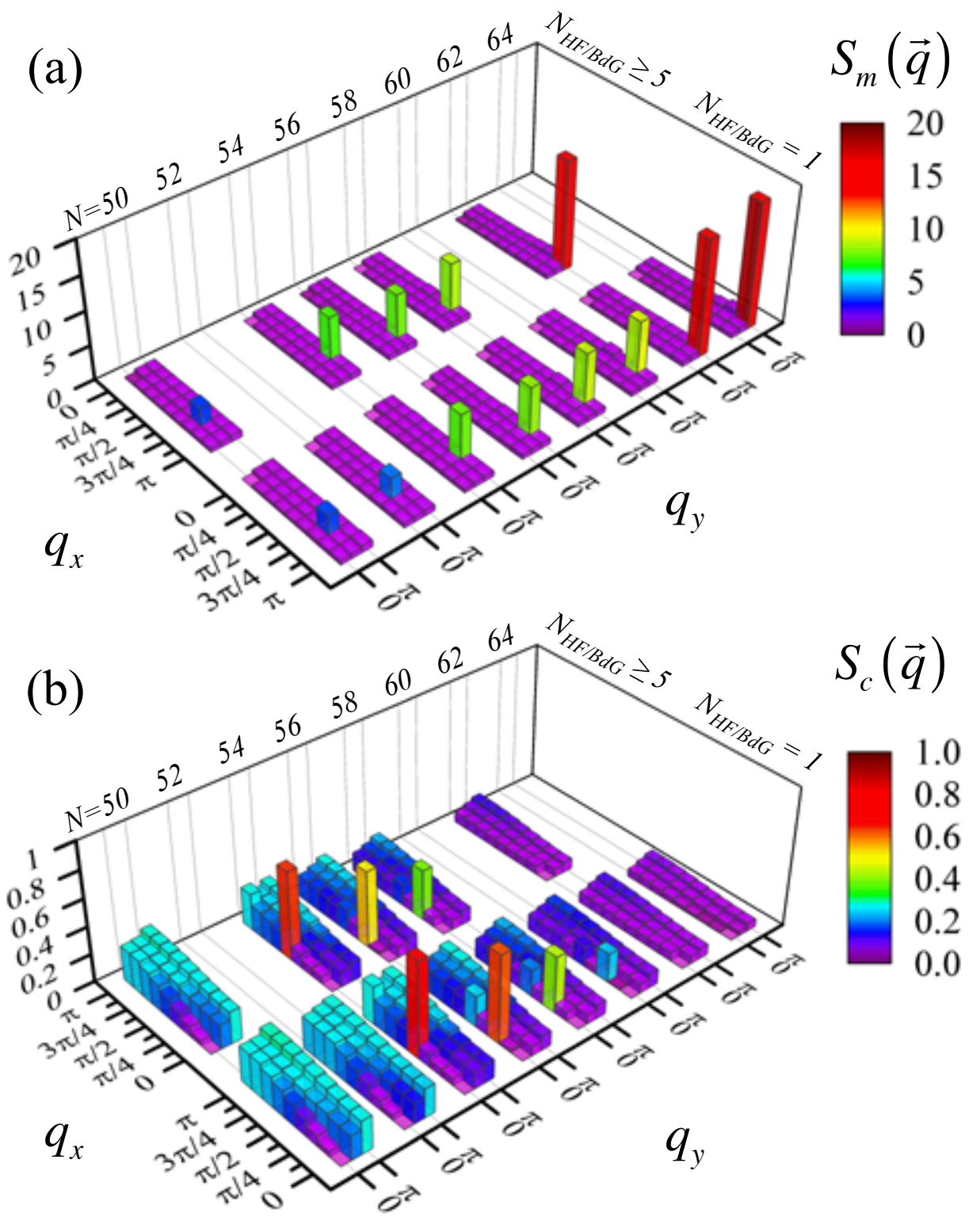

**Figure 8.** Momentum dependence of (a) magnetic and (b) charge structure factors for hole dopings $\delta < 1/4$ at large interaction strength $U/t = 12$. A rectangular $16 \times 4$ cell is considered. Spin and density autocorrelation functions are calculated from the numerical solution of the symmetry projected HF/BdG scheme. All symmetries are restored through projections on the number of atoms $N$, a zero total pseudo-momentum $\vec{K}$, the spin-singlet subspace and the irreducible representation $A_1$ of the $C_{2v}$ lattice symmetry group. The latter is physically associated to a many-body wavefunction invariant under horizontal and vertical mirrors. Note that these quantum number projection are also included during the energy minimization, except for the total spin where only its $z$-component and parity are imposed. In both parts (a) and (b), 3D-histograms in the front are obtained with one HF/BdG pair of states, while those in the back result from an enlarged subspace spanned by several sequentially optimized HF/BdG wavefunctions (five couples for $N = 50$, 54, 58, 62 and ten couples for $N = 56$. The cases $N = 52$, 60, 64 exhibit equivalent features regarding the number of HF/BdG pairs considered. They are not shown here for clarity's sake). Both magnetic and charge correlation functions show little sensitivity to the improvement of the variational state.

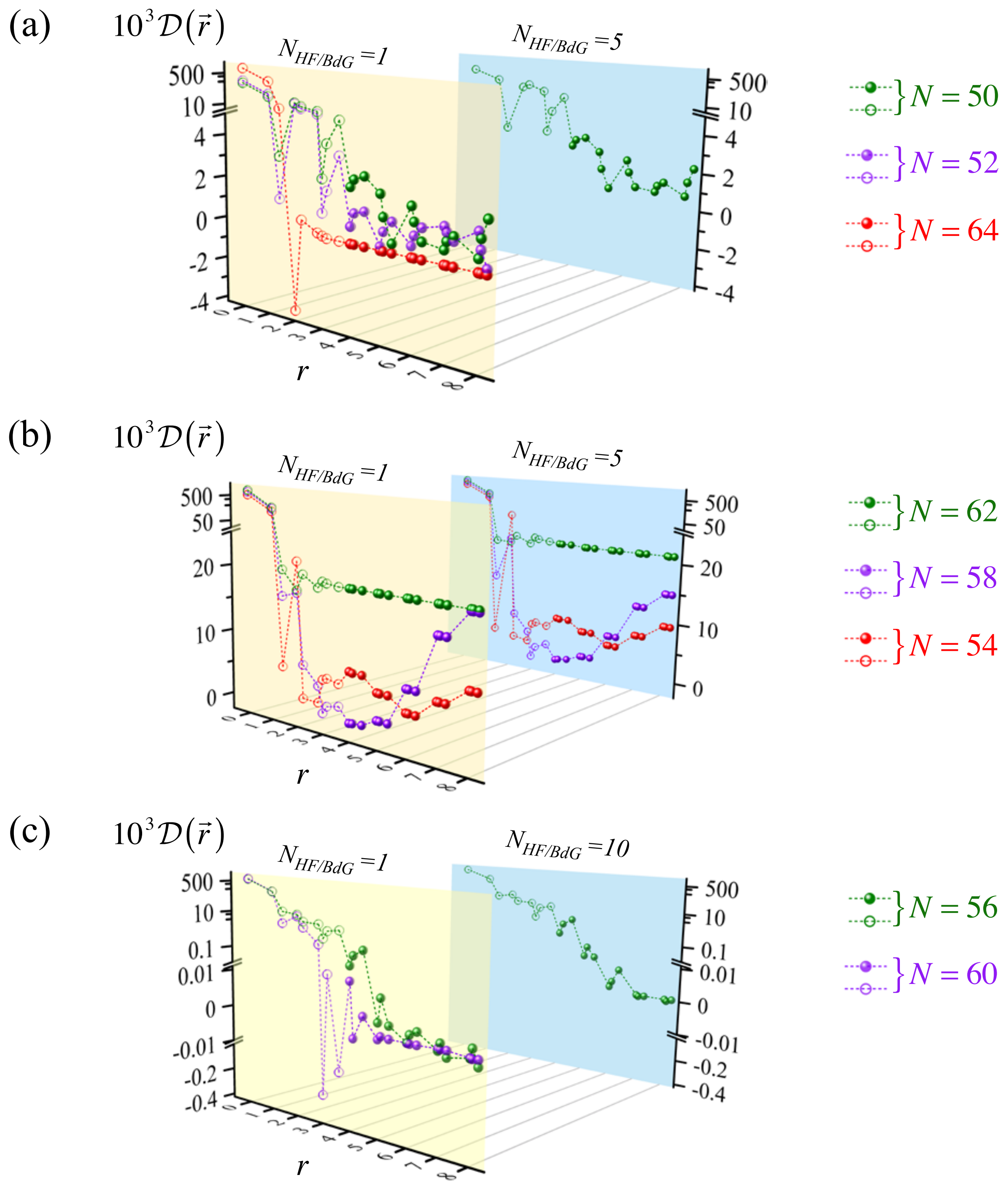

**Figure 9.** Dependence of the $d$-wave pair correlation function $\mathcal{D}(\vec{r})$ against separation distance $r$ for different numbers $N$ of atoms at strong coupling $U/t = 12$ [50]. In (b), note the oscillations at $r > 4$ in the stripe-like states with a full charge period $\lambda_c$ for $N = 54$ and half a period for $N = 58$. The same wavefunctions as for Fig. 8 are used. Short and long-ranged parts of $\mathcal{D}(\vec{r})$ are indicated by open and full symbols, respectively. They are not affected by the improvement of the variational ansatz, as shown in the back of parts (a), (b), and (c). As in Fig. 8, the cases $N = 52,\ 60,\ 64$ are omitted for clarity.

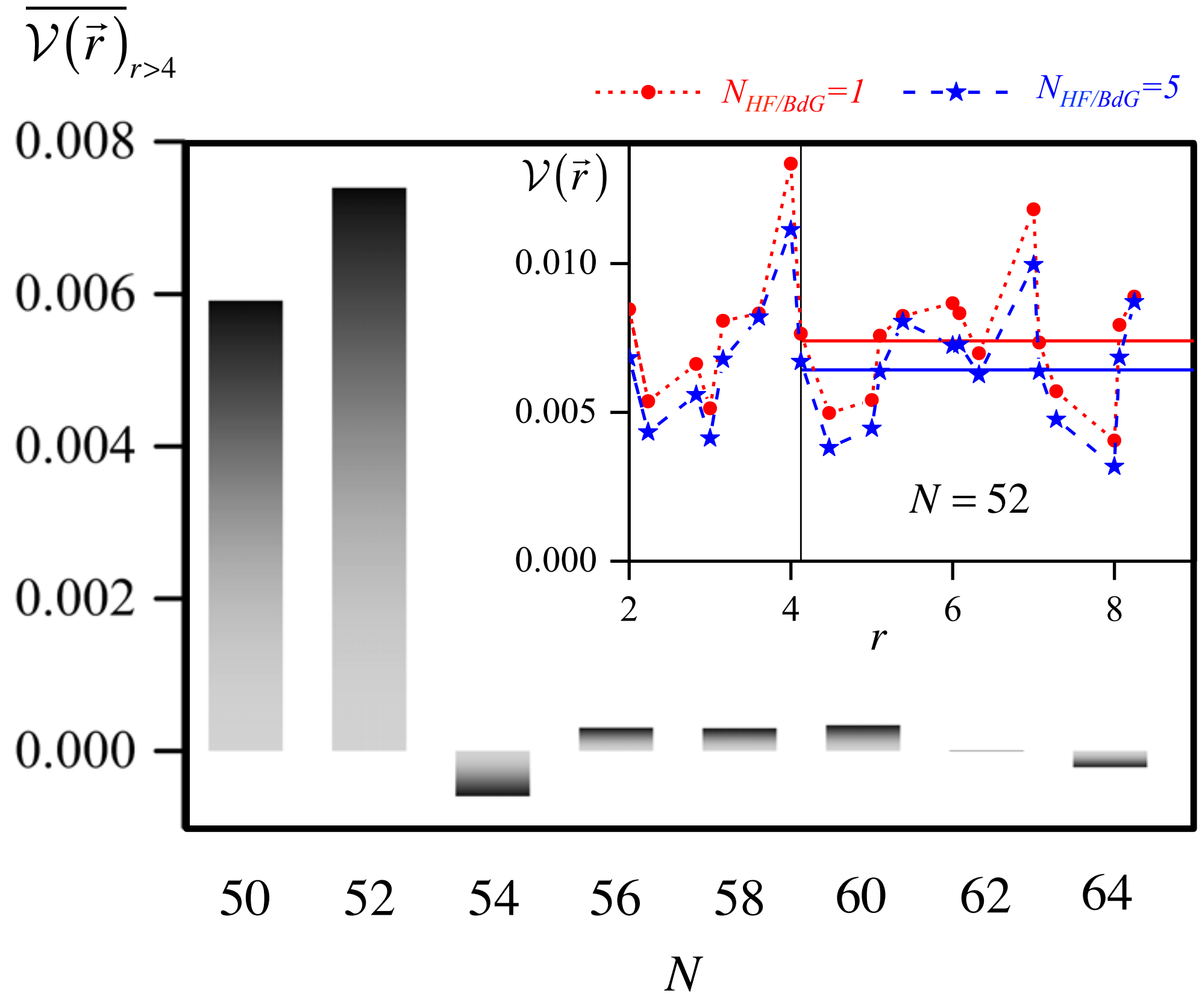

**Figure 10.** Spiral correlation function $\mathcal{V}(\vec{r})$ averaged over large distances $r > 4$ against the lattice filling for $U/t = 12$. The same wavefunctions as for Fig. 8 are used. However, performing the full spin projection for such a four-body observable is beyond reach. We limit ourselves here to impose the $z$-component $S_z$ and the spin-parity in addition to the restoration of all other symmetries. The detailed behavior of $\mathcal{V}(\vec{r})$ is shown in the inset for $N = 52$ atoms. Note the small difference between full circles and stars that correspond to one and five HF/BdG pair(s) of states, respectively.

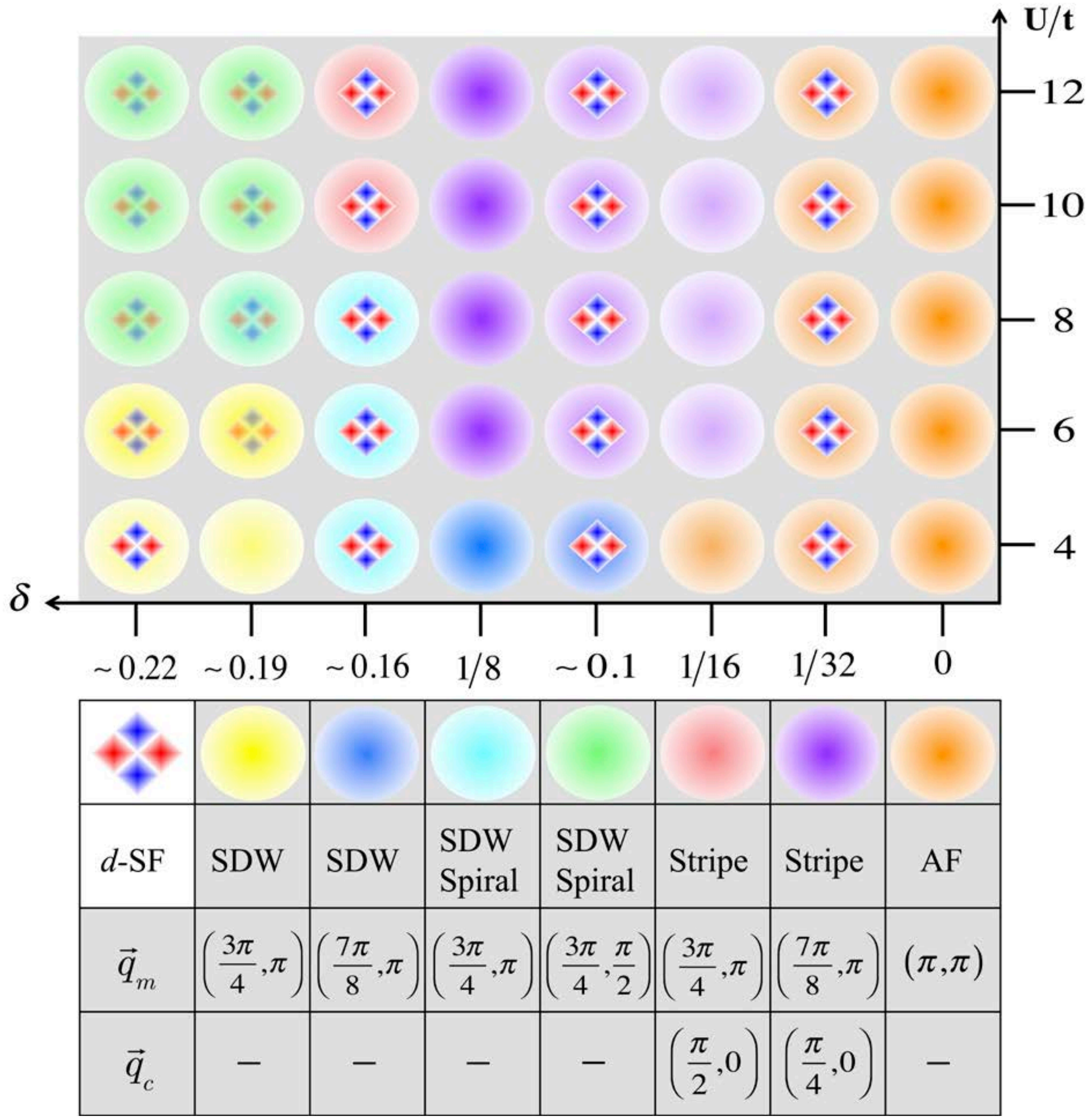

**Figure 11.** Phase diagram arising from the symmetry projected HF/BdG approach for repulsively interacting cold fermions loaded in optical four-leg tubes of length $L=16$. Colors refer to different magnetic (charge) orders revealed by a peak at the wavevector $\vec{q}_m$ ($\vec{q}_c$) in the Fourier transform of the spin (density) autocorrelation function. For each hole doping $\delta$ and interaction strength $U/t$, the $d$-wave superfluidity symbol is made more visible when the pair correlation function exhibits off-diagonal long-ranged order.

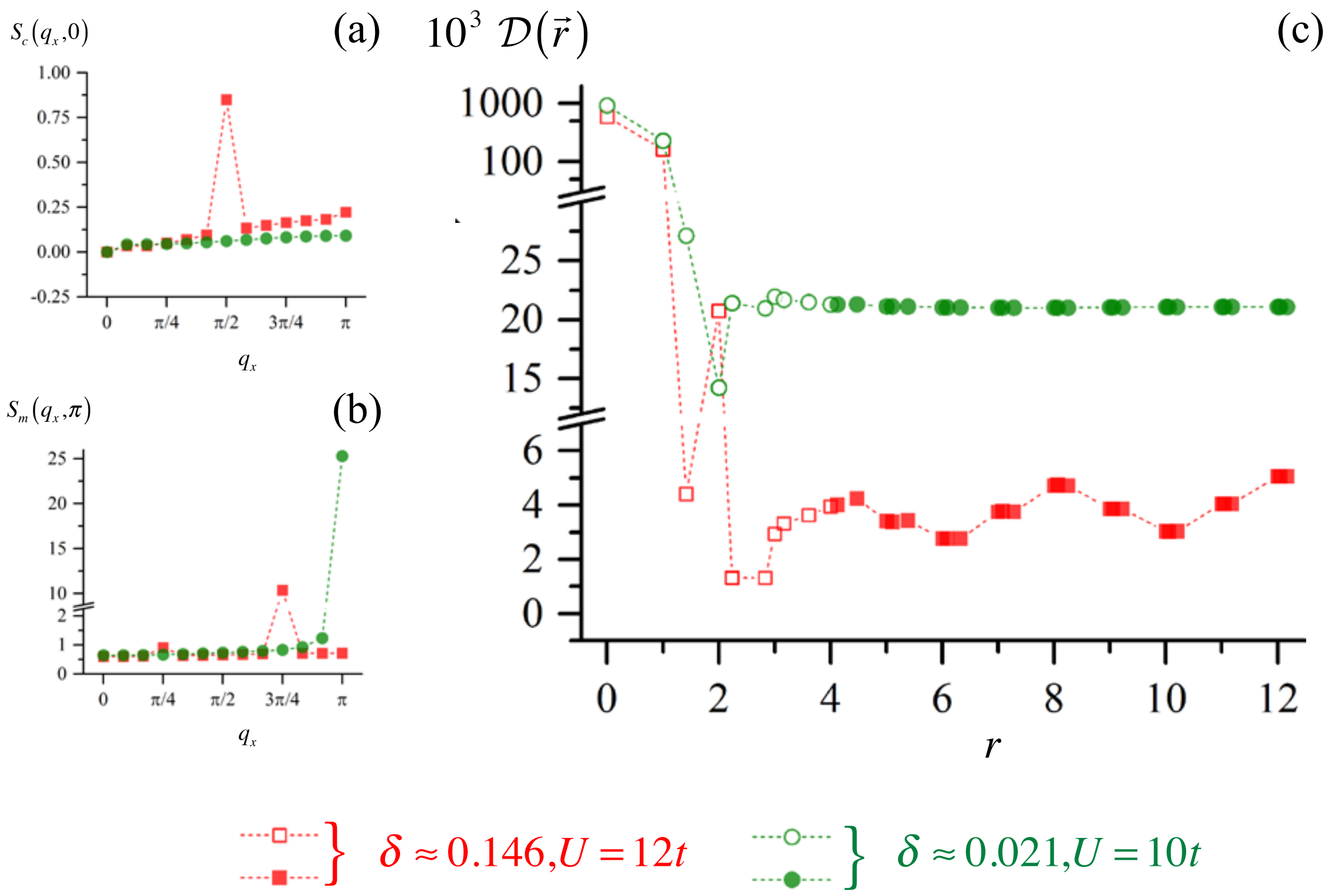

**Figure 12.** Charge (a) and spin (b) structure factors as a function of wavevectors along relevant paths of the first Brillouin zone. (c) Spatial dependence of the autocorrelation function $\mathcal{D}(\vec{r})$ in the $d$-wave pairing channel. Symmetry projected HF/BdG results are shown for a four-leg ladder of length $L = 24$.

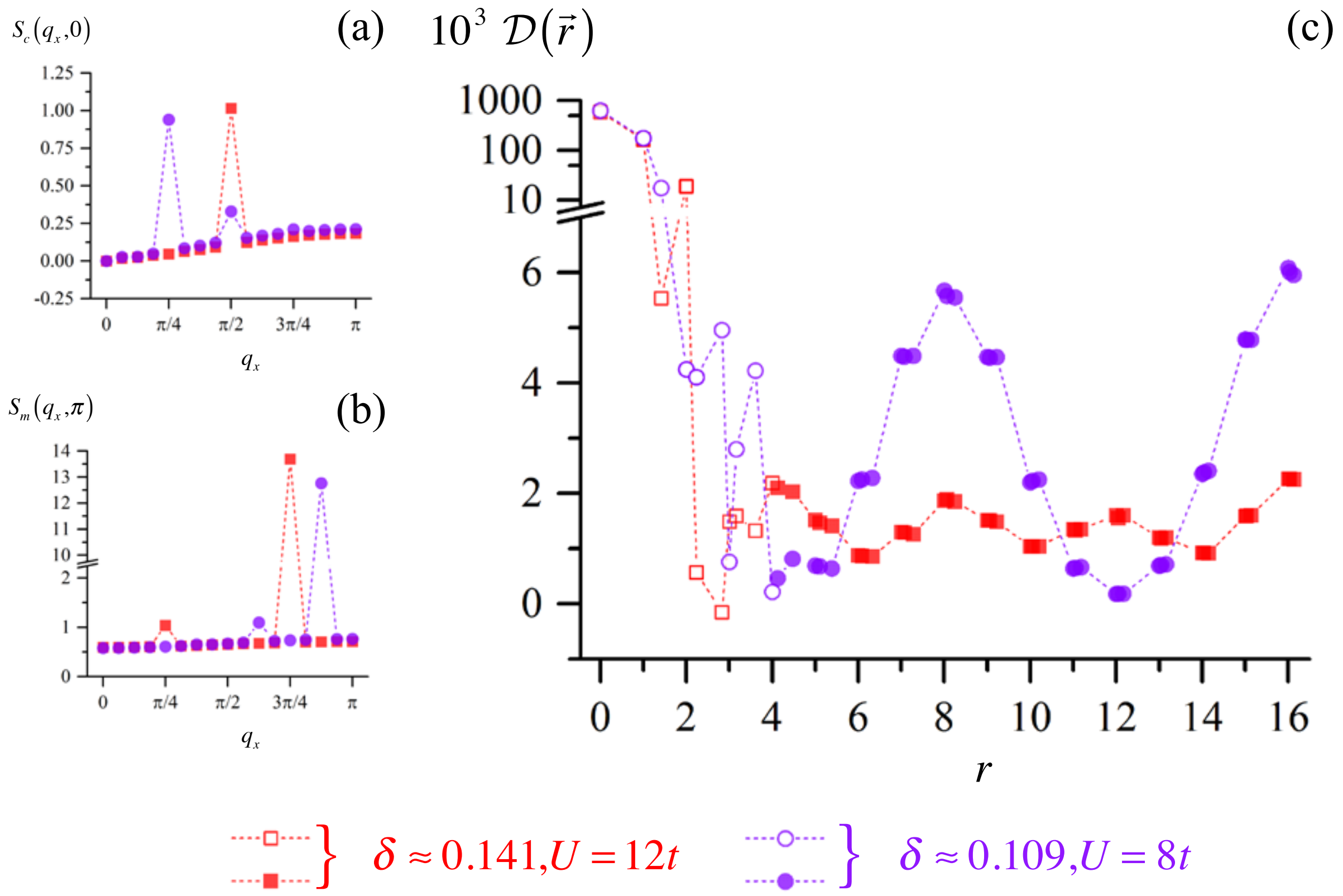

**Figure 13.** Same as in Fig. 12 but for a tube of $L = 32$ plaquettes and different lattice fillings.

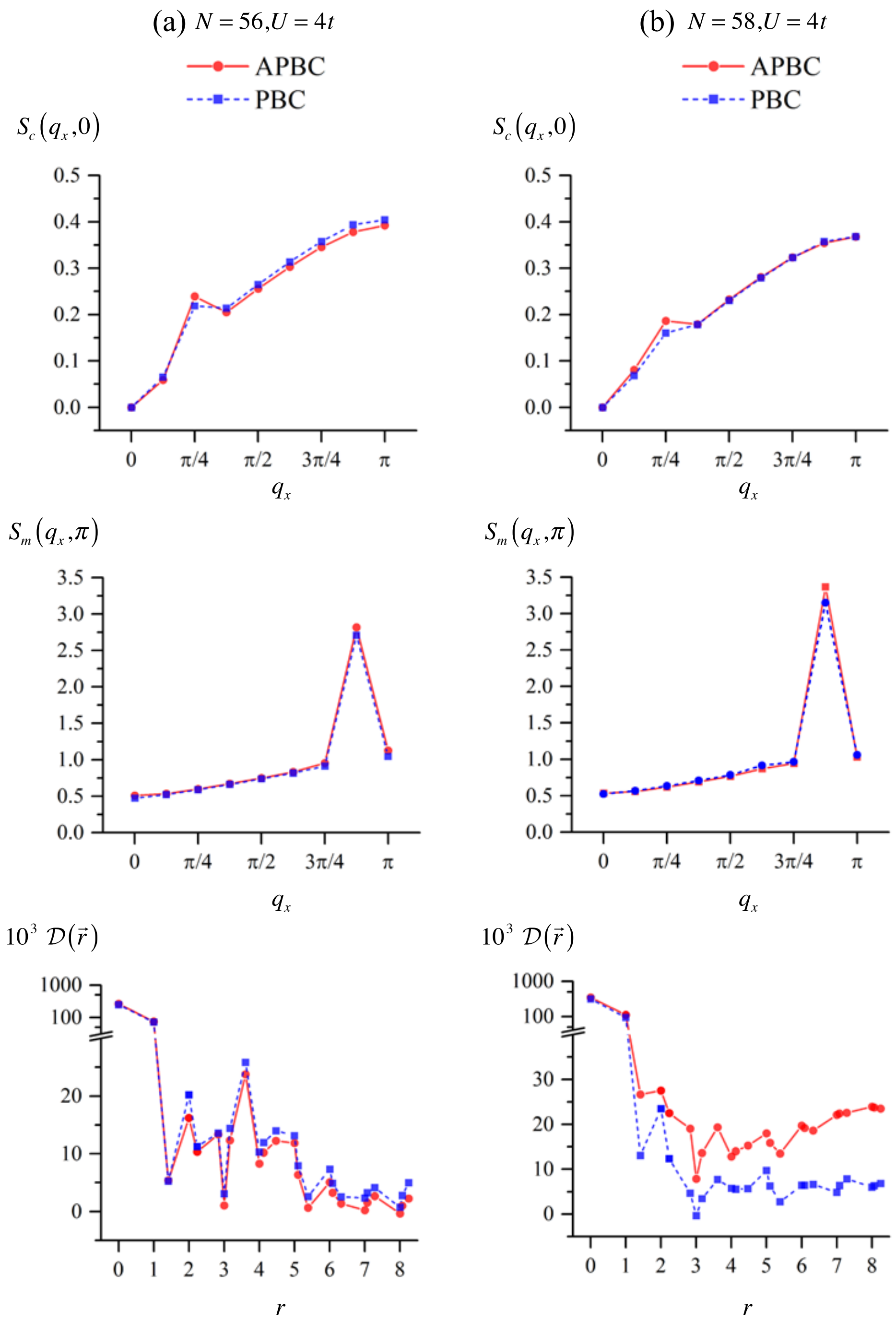

**Figure 14.** Sensitivity on boundary conditions along the leg direction of correlation functions in the density, spin and superfluid channels as obtained from the symmetry projected HF/BdG ansatz in a tube geometry of length $L=16$. The relevant charge $S_c(\vec{q})$ and magnetic $S_m(\vec{q})$ structure factors are shown on the upper and middle parts, respectively, while the *d*-wave pair correlation function $\mathcal{D}(\vec{r})$ is plotted at the bottom. The variational energies with APBC (PBC) are $E_\Gamma=-63.445\ t\ (-63.592\ t)$ and $E_\Gamma=-60.885\ t\ (-61.297\ t)$ for $N=56$ and $N=58$, respectively.

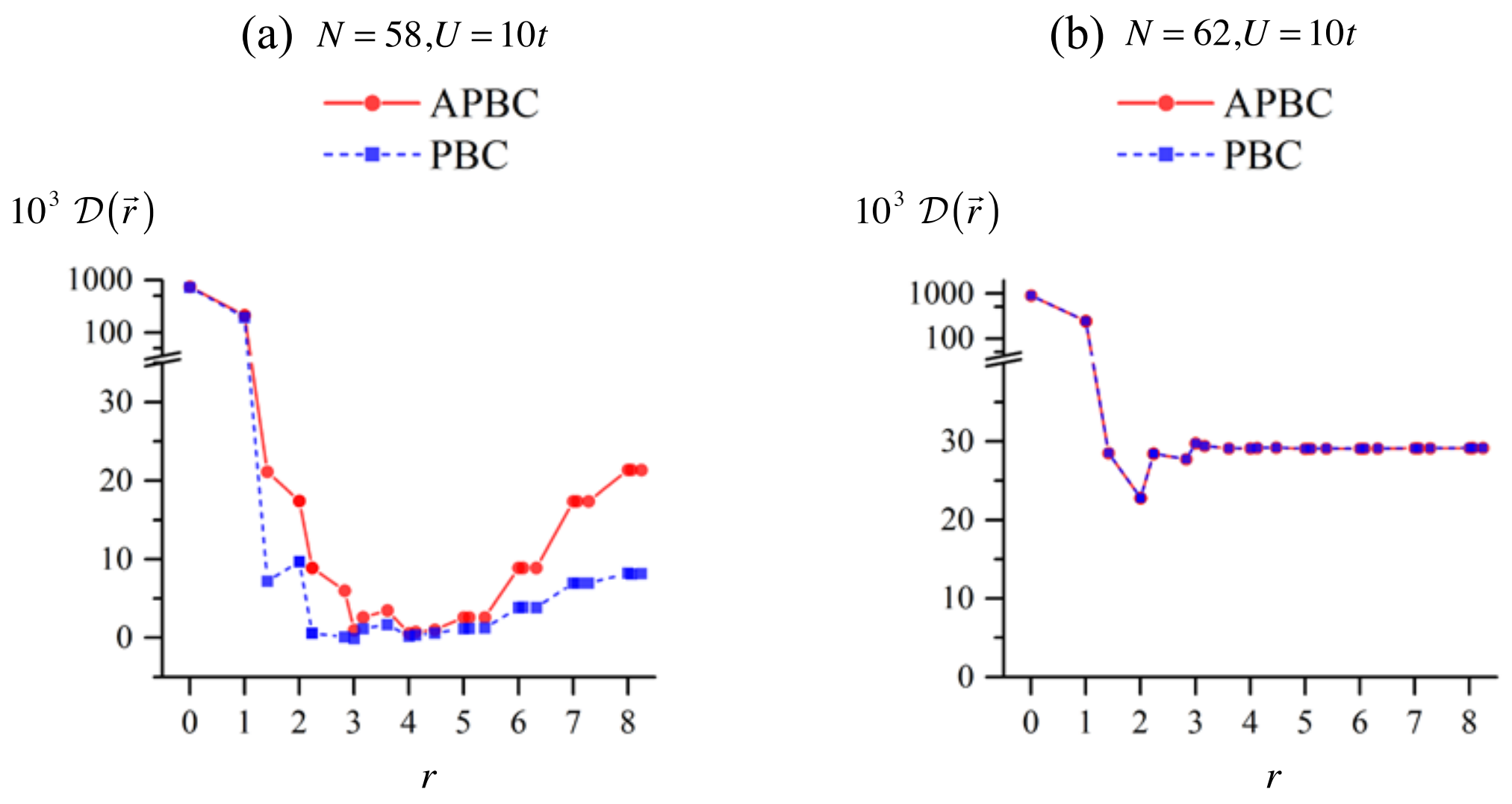

| | APBC | PBC |
|---|---|---|
| $E_\Gamma/t$ | -34.9927 | -35.259 |
| $\vec{q}_c$ | (π/4, 0) | (π/4, 0) |
| $S_c(\vec{q}_c)$ | 0.384 | 0.385 |
| $\vec{q}_m$ | (7π/8, π) | (7π/8, π) |
| $S_m(\vec{q}_m)$ | 5.666 | 5.664 |

| | APBC | PBC |
|---|---|---|
| $E_\Gamma/t$ | -29.3844 | -29.3846 |
| $\vec{q}_c$ | | |
| $S_c(\vec{q}_c)$ | | |
| $\vec{q}_m$ | (π, π) | (π, π) |
| $S_m(\vec{q}_m)$ | 16.3214 | 16.3215 |

**Figure 15.** Dependence on the boundary conditions of several observables for a ladder of $L = 16$ four-site plaquettes at two different filling factors. The long-ranged pairing correlations $\mathcal{D}(\vec{r})$ and the energy $E_\Gamma$ are lowered by PBC in the stripe phase $N = 58$, while this is not the case in the doped antiferromagnet $N = 62$. The location $\vec{q}_c$ ($\vec{q}_m$) and the values of the peaks in the charge (spin) structure factors $S_c$ ($S_m$) are not affected by the boundary conditions.